\documentclass[showpacs,preprintnumbers,amsmath,amssymb,aps, prd]{revtex4}
\usepackage{graphicx}
\usepackage{grffile}
\usepackage{mathrsfs, mathtools}
\usepackage{caption}
\usepackage{float}
\usepackage{xcolor}
\usepackage{changepage}
\usepackage{dcolumn}
\usepackage{bm}
\usepackage{graphicx,subfigure,epsfig}
\usepackage{multirow}
\newcommand{\lagr}{\mathscr{L}}

\def\f{\frac}
\def\rs{\rho_s}

\def\c{\cite}
\def\r{\ref}
\def\fr{f(r)}

\def\s{Schwarzschild }
\newcommand\be{\begin{equation}}
\newcommand\ee{\end{equation}}
\newcommand\ba{\begin{eqnarray}}
\newcommand\ea{\end{eqnarray}}
\newcommand\nn{\nonumber}
\newcommand\lt{\left}
\newcommand\rt{\right}
\newcommand\pt{\partial}

\newcommand\mc{\mathcal}
\begin{document}
\title{Shadow, ISCO, Quasinormal modes, Hawking spectrum, Weak Gravitational lensing, and parameter estimation of a Schwarzschild Black Hole Surrounded by a
Dehnen Type Dark Matter Halo}
\author{Sohan Kumar Jha}
\email{sohan00slg@gmail.com}
\affiliation{Chandernagore College, Chandernagore, Hooghly, West
Bengal, India}

\date{\today}
\begin{abstract}
\begin{center}
Abstract
\end{center}
We consider \s black hole (BH) embedded in a Dehnen-$(1,4,0)$ type dark matter halo (DDM) with two additional parameters - core radius $r_s$ and core density $\rs$ apart from mass $M$. We analyze the event horizon, photon orbits, and ISCO around DDM BHs and emphasize the impact of DDM parameters on them. Our study reveals that the presence of dark matter (DM) favourably impacts the radii of photon orbits, the innermost stable circular orbit (ISCO), and the event horizon. We find the expressions for specific energy and angular momentum for massive particles in time-like geodesics around DDM BH and investigate their dependence on DDM parameters. We display BH shadows for various values of core density and radius that reveal larger shadows cast by a \s BH surrounded by DDM (SDDM) than a \s BH in vacuum (SV). We then move on to study quasinormal modes (QNMs) with the help of the $6th$ order WKB method, the greybody factor using the semi-analytic bounds method, and the Hawking spectrum for scalar and electromagnetic perturbations. Core density and radius are found to have a significant impact on QNMs. Since QNMs for scalar and electromagnetic perturbations differ significantly, we can differentiate the two based on QNM observation. The greybody factor increases with core density and radius, whereas, the power emitted as Hawking radiation is adversely impacted by the presence of DM. We then study the weak gravitational lensing using the Gauss-Bonnet theorem and obtain the deflection angle with higher-order correction terms. Here, we see the deflection angle gets enhanced due to DM. Finally, we use bounds on the deviation from \s, $\delta$, reported by EHT for $M87^*$, Keck, and VLTI observatories for $Sgr A^*$ to gauge the viability of our model. Our model is found to be concordant with observations. This leads to the possibility of our galactic center being surrounded by DDM.
\end{abstract}
\maketitle
\section{Introduction}
The existence of DM is one of the most sought-after topics in physics. Our understanding of various astrophysical scenarios posits that BHs are surrounded by matter fields. DM qualifies as one of the plausible candidates for the matter field that may surround BHs. Observations related to giant elliptical and spiral galaxies have provided the first breakthrough in the search for DM \cite{rubin}. One recent study has pegged the contribution of DM in a galaxy's mass up to $90\%$ \c{persic}. There exists strong astrophysical evidence that points towards DM halos embedding supermassive BHs (SMBHs) that reside in galactic centers \cite{akiyamal1,akiyamal6}. This makes it imperative to reckon the contribution of DM near galactic center \c{sofue, boshkaye}. Different DM profiles can be considered to incorporate the effect of DM [\citenum{kiselev} - \citenum{rayimbaev}]. Dehnen density profile is commonly considered for dwarf galaxies \c{dehnen, mo}. This manuscript considers the static and spherically symmetric solution reported in \c{kalyan} where the Dehnen-$(1,4,0)$ type DM profile is considered. Particle motion around BHs encodes information not only of the intrinsic space-time geometry but also of the surrounding matter field. This makes it imperative to study the BH shadow and ISCO to apprehend the effect of DM. BH shadow is essentially a dark region in the celestial plane circumferenced by a bright emission ring, whereas ISCO is a stable orbit closest to a BH for a test particle. The dependence of BH shadow and ISCO on the BH parameters makes them a productive field of study. Please see [\citenum{RC}-\citenum{SN}] for more details regarding the application of BH shadow in the detection of DM and [\citenum{sanjar}-\citenum{sobhan}] for the effect of DM on ISCO. \\

Normal modes are characteristics of closed systems, whereas QNMs are related to open systems. When a BH is perturbed, after the initial outburst of radiation comes the ringdown phase where QNMs arise stemming from the dissipative oscillation of space-time \c{KONOR}. In the ringdown phase, oscillations do not depend on initial conditions. As such, QNMs are the sole function of intrinsic parameters of BH and the type of perturbation responsible for QNMs. During the oscillations, the whole system loses energy through the emission of gravitational waves (GWs). QNMs are complex-valued frequencies where the real part gives the frequency of the emitted GWs, and the imaginary part provides information regarding the stability of the underlying space-time: a negative value signifies stability against perturbation. They represent the decay rate of GWs. This makes the study of QNMs a potent tool not only to gauge the effect of DM but also to probe the stability of the BH-matter system. Studies related to QNMs for various BHs have been carried out in [\citenum{13}-\citenum{jha4}].\\

Hawking radiation arises due to the unification of GR and quantum mechanics. Hawking, with the help of quantum field theory, showed that BHs emit radiation as a consequence of particles escaping BH \c{HAWKING}. The emitted radiation, called Hawking radiation, is constituted by one of the particles created in a pair production near the event horizon that can escape BH gravitational field, and the second particle gets annihilated [\citenum{HH}-\citenum{HC}]. A temperature is associated with a BH  to be consistent with thermodynamics \c{BEK, KEIF}. Various techniques are available to obtain Hawking temperature [\citenum{SW}-\citenum{SI}]. The emission spectrum of a BH differs from a blackbody spectrum. This is because, in the background of a BH, particles move in a potential. While some particles get transmitted toward spatial infinity, the rest are reflected back toward BH and consumed. The greybody factor is associated with the quantum nature of a BH that provides the transmission probability of radiation and corrects the blackbody spectrum. Different strategies are there which provide greybody factor [\citenum{qn32}-\citenum{WJ}]. Here, we employ semi-analytic bounds method \c{GB, GB1, GB2}.\\

When a light ray passes near a massive object such as SMBH, its path bends. This way, gravitational lensing provides an excellent avenue to detect massive astronomical objects. GL is also employed to probe the Universe's expansion. A significant number of studies have been devoted to studying GL for BHs, naked singularities, and wormholes [\citenum{keeton}-\citenum{511}]. This article deals with GL in the weak field limit known as weak lensing. We use the Gauss-Bonnet theorem first proposed by Gibbons and Werner in \c{GW}. It was later extended to various cases \c{WERNER, ISHIHARA1, ISHIHARA2, ONO1, ONO2, ONO3}. We obtain higher order correction terms in the deflection angle using \cite{CRISNEJO}. Theoretical predictions of multifarious astrophysical phenomena provide information regarding BH and its surrounding environment, whereas experimental observations present an outstanding opportunity to probe the efficacy of our models. Observations related to the shadow of $M87^*$ and $Sgr A^*$ reported by the Event Horizon Telescope, Keck, and VLTI observatories provide an unparalleled opportunity to put our model to the test. We use bounds on the deviation from \s for $M87^*$ \c{M871, M872} and $Sgr A^*$ \c{keck, vlti1, vlti2} to constrain DM parameters.\\

We organize our article as follows: Section II is where we introduce the BH solution and discuss the impact of DM on the event horizon. We discuss QNMs in section III, and section IV is devoted to studying the greybody factor and Hawking spectrum. Section V deals with weak gravitational lensing and we try to constrain core density and radius in section VI. This article ends with conclusions in section VII. We have used $G=c=M=1$ throughout the paper.
\section{BH surrounded by Dehnen-type DM}
Authors in article \c{kalyan} have reported a static and spherically symmetric metric where the \s BH is embedded in a Dehnen-$(1,4,0)$ type DM halo. The metric is
\be
ds^2=-\fr dt^2+\f{dr^2}{\fr}+r^2\lt(d\theta^2+\sin^2 \theta d\phi^2\rt),\label{metric}
\ee
where $\fr=1-\frac{2 M}{r}-\frac{4 \pi  \left(r_s+2 r\right) r_s^3 \rho _s}{3 \left(r_s+r\right){}^2}$, $\rs$ and $r_s$ being core density and core radius respectively. The above solution reduces to a \s metric in vacuum in the limit $\rs \rightarrow 0$ or $r_s \rightarrow 0$. The metric [\r{metric}] has two singularities: one at $r=0$ and another where $\fr=0$. Kretschmann scalar for the metric [\r{metric}] is
\ba \nn
&&K=\\\nn
&&\frac{16}{9 r^6 \left(r_s+r\right){}^8} \left(4 r^6 r_s^2 \left(189 M^2+192 \pi  M r_s^3 \rho _s+46 \pi ^2 r_s^6 \rho _s^2\right)+4 r^5 r_s^3 \left(378 M^2+231 \pi  M r_s^3 \rho _s+44 \pi ^2 r_s^6 \rho
   _s^2\right)\right.\\\nn
&&\left. +2 r^4 r_s^4 \left(945 M^2+348 \pi  M r_s^3 \rho _s+52 \pi ^2 r_s^6 \rho _s^2\right)+8 r^3 r_s^5 \left(189 M^2+42 \pi  M r_s^3 \rho _s+4 \pi ^2 r_s^6 \rho
   _s^2\right)\right.\\\nn
&&\left. +4 r^2 r_s^6 \left(189 M^2+24 \pi  M r_s^3 \rho _s+\pi ^2 r_s^6 \rho _s^2\right)+27 M^2 r_s^8+3 r^8 \left(3 M+4 \pi  r_s^3 \rho _s\right){}^2+24 r^7 r_s \left(3
   M+\pi  r_s^3 \rho _s\right)\right.\\\nn
&&\left. \left(3 M+4 \pi  r_s^3 \rho _s\right)+12 M r r_s^7 \left(18 M+\pi  r_s^3 \rho _s\right)\right).
\ea
The Kretschmann scalar has one singularity at $r=0$. Thus, the singularity of the BH solution [\r{metric}] at $r=0$ is an essential singularity that cannot be removed by any coordinate transformation, whereas, the singularity at $\fr=0$ is a removable singularity. The largest solution of the equation $\fr=0$ provides the position of the event horizon $r_h$. However, this equation for the metric under consideration has no analytical solution. Thus, we will resort to a numerical solution to obtain the event horizon. Qualitative variations of the event horizon with core radius and density are shown below.
\begin{figure}[H]
\begin{center}
\begin{tabular}{cc}
\includegraphics[width=0.4\columnwidth]{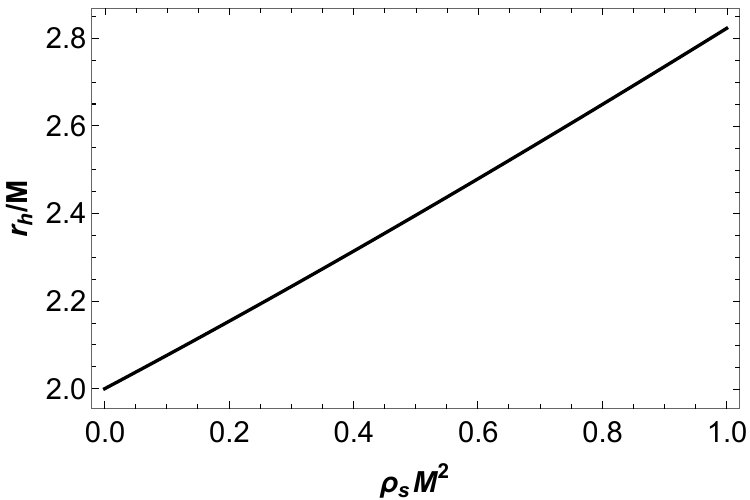}&
\includegraphics[width=0.4\columnwidth]{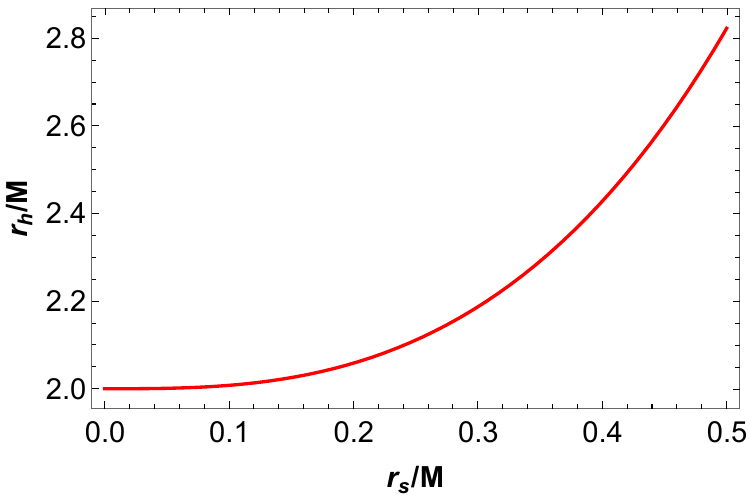}
\end{tabular}
\caption{Variation of event horizon with $\rs$ kepping $r_s=0.5M$ shown in left panel and with $r_s$ kepping $\rs M^{2}=1.0$ shown in the right panel. }\label{rh}
\end{center}
\end{figure}
Fig. [\r{rh}] shows that the event horizon increases linearly with core density for a fixed value core radius. On the other hand, for a fixed value of core density, $r_h \propto r_s^{3}$. Next, we will consider null and time-like geodesics to study the impact of DM on shadow and ISCO.
\section{Shadow radius and inner-most circular orbit}
We, in this section, consider null and time-like orbits on the equatorial plan, i.e., $\theta=\f{\pi}{2}$. The Lagrangian in the background of the BH under consideration is
\be
\lagr=\f{1}{2}(-\fr \dot{t}^2+\f{\dot{r}^2}{\fr}+r^2 \dot{\phi}^2),
\ee
where over-dot is differentiation with respect to affine parameter $\tau$. Since the metric is independent of $t$ and $\phi$, we have two conserved quantities of motion - energy $E$ and angular momentum $L$. For mass-less particles, $E$ and $L$ are total energy and angular momentum, respectively, whereas, for massive particles, they are specific energy and angular momentum. They are given by
\be
E=-\f{\pt \lagr}{\pt \dot{t}}=\fr \dot{t} \quad \text{and} \quad L=\f{\pt \lagr}{\pt \dot{\phi}}=r^2 \dot{\phi}
\ee
Four-velocity of particles (massive and mass-less) follow the relation $\dot{x_\mu}\dot{x^\mu}=\epsilon$ where $\epsilon$ is zero for mass-less particles and $-1$ for massive particles. With this, we obtain the following equation for radial coordinates:
\be
\dot{r}^2=E^2 - \fr \lt(-\epsilon+\f{L^2}{r^2}\rt)=E^2 - V(r),
\ee
where $V(r)$ is the potential for the particle. Circular orbits are attributed by $\dot{r}=\ddot{r}=0$. $V(r)=E^2$ follows from the first condition whereas the second condition yields $\f{dV(r)}{dr}=0$. From the second condition, we obtain the radius of the photon orbit $(r_{ph})$ as the solution of the equation
\be
r_{ph}f'(r_{ph})=2f(r_ph),
\ee
where $'$ in the above equation is differentiation with respect to $r$. The critical impact parameter $b_{ph}$ corresponding to the photon orbit, which is also the shadow radius $(R_s)$ of the BH, is
\be
b_{ph}=R_s=\f{L}{E}=\f{r_{ph}}{\sqrt{f(r_{ph})}.}
\ee
It is not possible to obtain analytical expressions for either photon radius or critical impact parameter. We will, therefore, obtain their values numerically. We display variation of photon radius and shadow radius with core density and radius in Fig. [\r{rsrp}]. It shows that both the radii increase with core density as well as radius. Both the radii increase linearly with the core density. However, the change in the radii with core radius is proportional to $r_s^3$. \\
\begin{figure}[H]
\begin{center}
\begin{tabular}{cc}
\includegraphics[width=0.4\columnwidth]{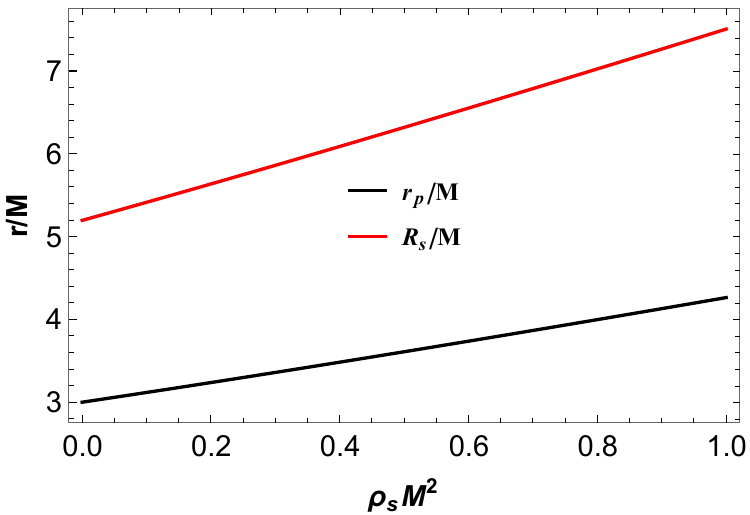}&
\includegraphics[width=0.4\columnwidth]{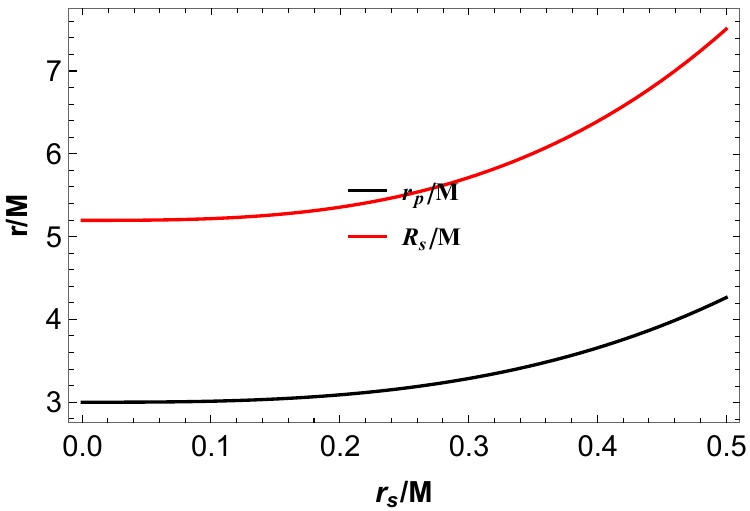}
\end{tabular}
\caption{Variation of photon and shadow radii with $\rs$ kepping $r_s=0.5M$ shown in left panel and with $r_s$ kepping $\rs M^{2}=1.0$ shown in the right panel. }\label{rsrp}
\end{center}
\end{figure}
We display the variation of shadow with core radius and density. It is evident from Fig. [\r{shadow}] that the size of the shadow increases due to the presence of DM.\\
\begin{figure}[H]
\begin{center}
\begin{tabular}{cc}
\includegraphics[width=0.4\columnwidth]{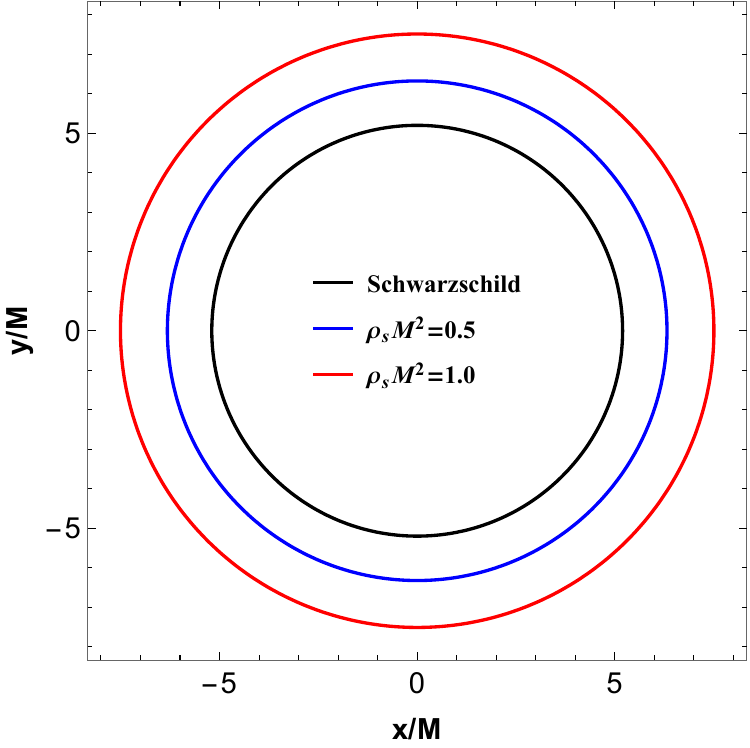}&
\includegraphics[width=0.4\columnwidth]{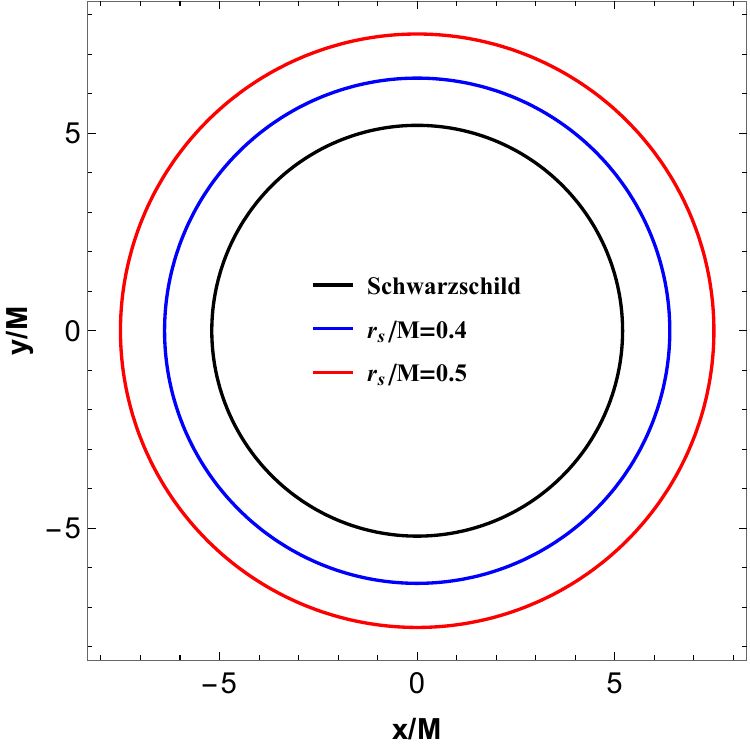}
\end{tabular}
\caption{BH shadows for different values of $\rs$ kepping $r_s=0.5M$ shown in left panel and for different values of $r_s$ kepping $\rs M^{2}=1.0$ shown in the right panel. }\label{shadow}
\end{center}
\end{figure}
Next, we turn our attention to time-like geodesics where two conditions $\dot{r}=\ddot{r}=0$ provide the following expressions for specific energy and angular momentum:
\ba\nn
&&E^2\\\nn
&=&\f{2\fr^2}{2\fr-rf'(r)}\\
&=&\frac{2 \left(\frac{2 M}{r}+\frac{4 \pi  \left(r_s+2 r\right) r_s^3 \rho _s}{3 \left(r_s+r\right){}^2}-1\right){}^2}{-\frac{6 M}{r}-\frac{8 \pi  \left(3 r^2+3 r r_s+r_s^2\right)
   r_s^3 \rho _s}{3 \left(r_s+r\right){}^3}+2},\\\nn
&&L^2\\
&=&\f{r^3 f'(r)}{2\fr-rf'(r)}\\\nn
&=&\frac{r^2 \left(r_s^3 \left(3 M+4 \pi  r^3 \rho _s\right)+3 M r^3+9 M r^2 r_s+9 M r r_s^2\right)}{-3 r_s^3 \left(3 M+4 \pi  r^3 \rho _s-r\right)+3 r^3 (r-3 M)+9 r^2 (r-3 M)
   r_s+9 r (r-3 M) r_s^2-12 \pi  r^2 r_s^4 \rho _s-4 \pi  r r_s^5 \rho _s}.
\ea
Fig. [\r{e}] demonstrates the behavior of $E$ vs $r/M$ for various values of core radius and density. It shows that the specific energy of a test particle in the background of a \s BH surrounded by Dehnen-type DM is always less than that for a \s BH in vacuum beyond ISCO radius. The minimum of the specific energy occurs at the ISCO radius. It is also observed that the specific energy tends to have a constant value far from BH. The impact of DM on the specific angular momentum is illustrated in Fig. [\r{l}], where a comparison is also made with \s BH in vacuum. It is evident from Figs. [\r{la}, \r{lb}] that both the parameters adversely impact the specific angular momentum.\\
\begin{figure}[H]
\centering
\subfigure[]{
\label{ea}
\includegraphics[width=0.4\columnwidth]{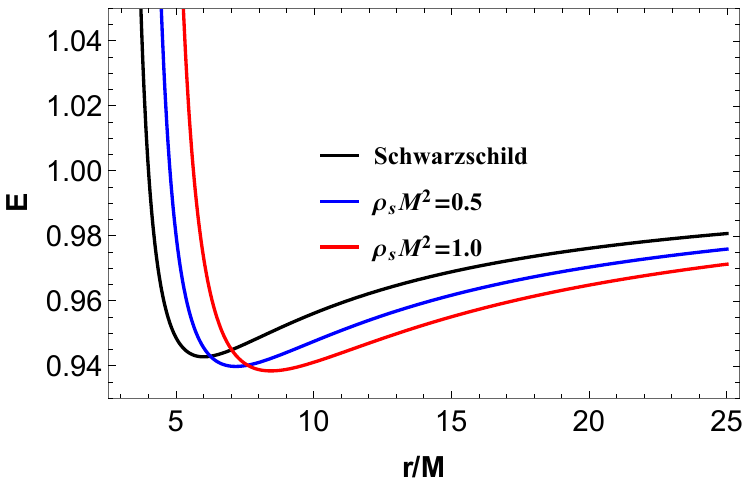}
}
\subfigure[]{
\label{eb}
\includegraphics[width=0.4\columnwidth]{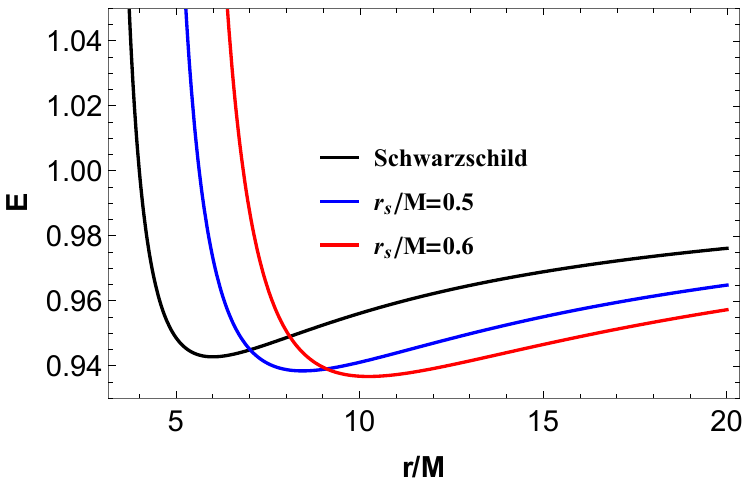}
}
\caption{Variation of $E$ with $r/M$ for different values of core density keeping $r_s=0.5M$ shown in the left panel and for different values of core radius keeping $\rs=1.0/M^2$ shown in the right panel. }
\label{e}
\end{figure}
\begin{figure}[H]
\centering
\subfigure[]{
\label{la}
\includegraphics[width=0.4\columnwidth]{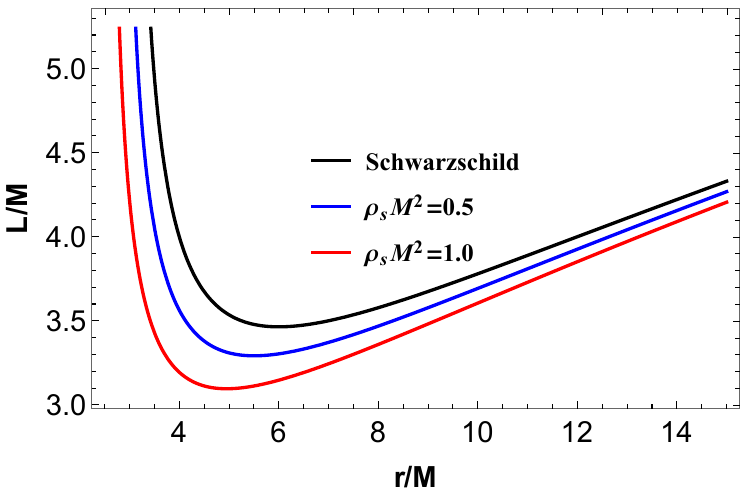}
}
\subfigure[]{
\label{lb}
\includegraphics[width=0.4\columnwidth]{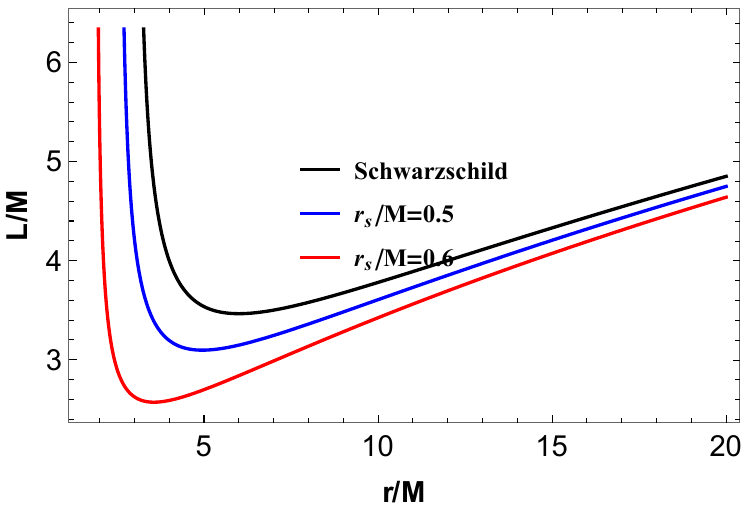}
}
\caption{Variation of $L$ with $r/M$ for different values of core density keeping $r_s=0.5M$ shown in the left panel and for different values of core radius keeping $\rs=1.0/M^2$ shown in the right panel. }
\label{l}
\end{figure}
An additional condition to be imposed on the potential for the inner-most circular orbit is $\f{d^2V(r)}{dr^2}|_{r=r_{isco}}=0$. This yields the following equation whose solution gives the ISCO radius:
\ba\nn
&&r_{isco}f(r_{isco}) f''(r_{isco})-2r_{isco}f'(r_{isco})^2+3f(r_{isco}) f'(r_{isco})=0\\\nn
&&\text{which gives}\\\nn
&&2 \left(3 M \left(r_s+r\right){}^4-2 \pi  r^3 r_s^3 \left(r_s-2 r\right) \rho _s\right) \left(6 M \left(r_s+r\right){}^2+4 \pi  r \left(r_s+2 r\right) r_s^3 \rho _s-3 r\left(r_s+r\right){}^2\right)\\\nn
&&+3 \left(r_s+r\right) \left(6 M \left(r_s+r\right){}^2+4 \pi  r \left(r_s+2 r\right) r_s^3 \rho _s-3 r \left(r_s+r\right){}^2\right) \left(-3 M
   \left(r_s+r\right){}^3+4 \pi  r^2 \left(r_s+r\right) r_s^3 \rho _s-4 \pi  r^2 \left(r_s+2 r\right) r_s^3 \rho _s\right)\\\nn
&&-4 \left(r_s^3 \left(3 M+4 \pi  r^3 \rho _s\right)+3 M r^3+9 M r^2 r_s+9 M r r_s^2\right){}^2=0.\\
\ea
The above equation does not have any analytical solution. As such, it is numerically solved to study the impact of DM on the motion of massive particles. We elucidate the impact of DM parameters on the ISCO radius graphically in Fig. [\r{isco}], which shows that similar to the cases of event horizon, photon, and shadow radii, the ISCO radius increases with core radius and density.The nature of dependence of ISCO radius on core density and radius is also similar to photon and shadow radii.
\begin{figure}[H]
\centering
\subfigure[]{
\label{}
\includegraphics[width=0.4\columnwidth]{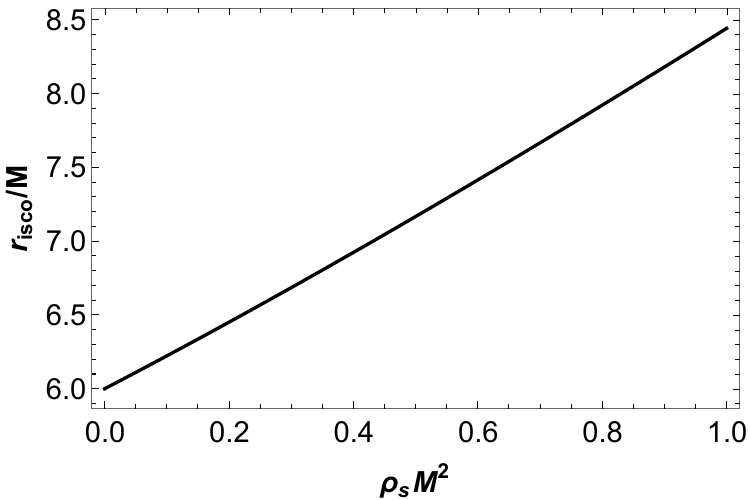}
}
\subfigure[]{
\label{}
\includegraphics[width=0.4\columnwidth]{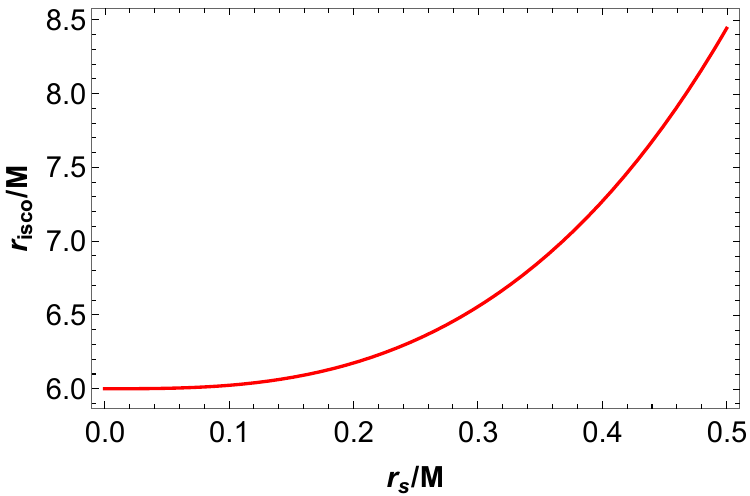}
}
\caption{Variation of ISCO radius with core density kepping $r_s=0.5M$ shown in left panel and with core radius kepping $\rs=1.0/M^2$ shown in the right panel.}
\label{isco}
\end{figure}
While Figs. [\r{rh} - \r{isco}] provide a qualitative view of the impact of DM parameters on various aspects of null and time-like geodesics, quantitative view will provide deeper insight into the effect of DM on the motion of mass-less and massive particles. Table [\r{rsrpa}] enlists numerical values of various quantities for different values of core density, whereas Table [\r{rsrpb}] enlists those values for different core radius.

\begin{center}
\begin{tabular}{|c|c|c|c|c|c|c|}
\hline
$\rho _s M^2$  & $r_h\text{/M}$ & $r_p\text{/M}$ & $R_s\text{/M}$ & $r_{\text{isco}}\text{/M}$ & $E_{\text{isco}}$ & $L_{\text{isco}}\text{/M}$ \\
\hline
 0. & 2. & 3. & 5.19615 & 6. & 0.942809 & 3.4641 \\
\hline
 0.2 & 2.15397 & 3.23735 & 5.63619 & 6.45115 & 0.941363 & 3.41469 \\
\hline
 0.4 & 2.31384 & 3.48337 & 6.08886 & 6.92373 & 0.940284 & 3.39079 \\
\hline
 0.6 & 2.47898 & 3.73702 & 6.55243 & 7.41476 & 0.939485 & 3.38547 \\
\hline
 0.8 & 2.64878 & 3.99735 & 7.02539 & 7.92164 & 0.9389 & 3.39393 \\
\hline
 1.0 & 2.82268 & 4.26351 & 7.50642 & 8.44215 & 0.938478 & 3.41272 \\
\hline
\end{tabular}
\captionof{table}{Numerical values of various quatities with several values of $\rs$ keeping $r_s=0.5M$.} \label{rsrpa}
\end{center}

\begin{center}
\begin{tabular}{|c|c|c|c|c|c|c|}
\hline
$r_s \text{/M}$  & $r_h\text{/M}$ & $r_p\text{/M}$ & $R_s\text{/M}$ & $r_{\text{isco}}\text{/M}$ & $E_{\text{isco}}$ & $L_{\text{isco}}\text{/M}$ \\
\hline
 0. & 2. & 3. & 5.19615 & 6. & 0.942809 & 3.4641 \\
\hline
 0.1 & 2.00779 & 3.01178 & 5.21688 & 6.02334 & 0.942793 & 3.46355 \\
\hline
 0.2 & 2.05838 & 3.08879 & 5.35472 & 6.17428 & 0.942581 & 3.45682 \\
\hline
 0.3 & 2.18691 & 3.28562 & 5.71225 & 6.55581 & 0.941825 & 3.43792 \\
\hline
 0.4 & 2.42776 & 3.65578 & 6.39065 & 7.26925 & 0.940361 & 3.41518 \\
\hline
 0.5 & 2.82268 & 4.26351 & 7.50642 & 8.44215 & 0.938478 & 3.41272 \\
\hline
\end{tabular}
\captionof{table}{Numerical values of various quatities with several values of $r_s$ keeping $\rs M^{2}=1.0$.} \label{rsrpb}
\end{center}
 Table [\r{rsrpa}] and [\ref{rsrpb}] show significant impact of DM on observabes regarding null and time-like geodesics. This concludes our discussion on the null and time-like geodesics in the background of the BH surrounded by Dehnen-type DM.
\section{Quasinormal frequencies}
\begin{figure}[H]
\centering
\subfigure[]{
\label{imb1}
\includegraphics[width=0.4\columnwidth]{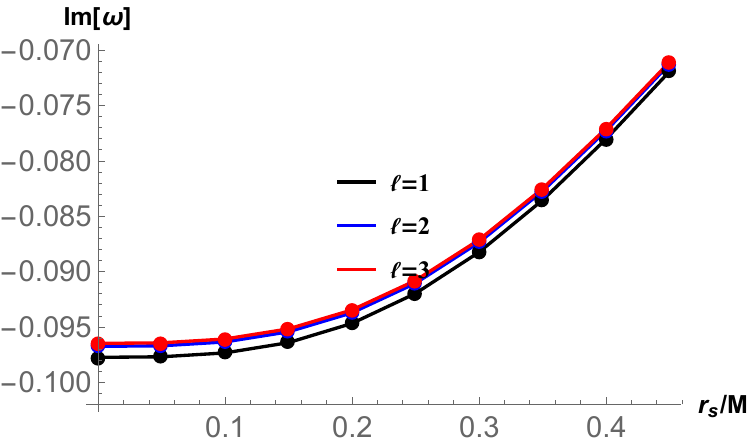}
}
\subfigure[]{
\label{imb2}
\includegraphics[width=0.4\columnwidth]{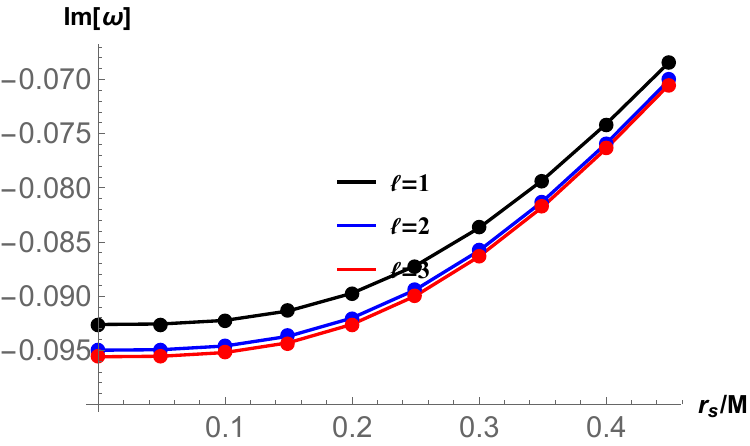}
}
\caption{Variation of decay rate with core radius for different $\ell$ keeping $\rs=1.0/M^2$. The left one is for scalar perturbation, and the right one is for electromagnetic perturbation.}
\label{imb}
\end{figure}

QNMs encode the inherent characteristics of the underlying geometry and are independent of any specific initial conditions. They also provide information regarding the stability of the black hole system against perturbations. To compute QNMs, we first take into account field equations given by
\ba
&&\frac{1}{\sqrt{-g}}{\partial \mu}(\sqrt{-g}g^{\mu\nu} \partial_{\nu}\chi) =0,\\\label{scalar}
&&\frac{1}{\sqrt{-g}}{\partial \nu }(F_{\rho\sigma}g^{\rho\mu}g^{\sigma\nu}\sqrt{-g})=0,\label{em}
\ea
where $ F_{\rho\sigma}={\partial \rho}A^\sigma-{\partial \sigma}A^\rho $, $A_\nu$ being the electromagnetic four-potential. Owing to the spherical symmetry in the existing system, we can decompose the field with the introduction of the tortoise coordinate defined by $\text{d}r_{\ast}=\f{\text{d}r}{\fr }$. This enables us to get a Schr$\ddot{o}$dinger-like equation
\be
-\frac{\partial^{2} \Phi}{\partial t^{2}}+\frac{\partial^{2} \Phi}{\partial r^{*2}} + V_{eff}(r^{*})\Phi = 0,\label{sch1}
\ee
where $V_{eff}(r)$ is the effective potential or the \textit{Regge--Wheeler} potential given by
\ba
V_{eff}(r)&=&\fr\left(\frac{\ell\left( \ell+1\right) }{r^{2}}+\fr^{\prime }\frac{(1-s^{2})}{r}\right)\\\nn
&= & \lt(1-\frac{2 M}{r}+\frac{4 \pi  \left(2 r_s+3 r\right) r_s^4 \rho _s}{3 r \left(r_s+r\right){}^2}\rt)\\
&&\lt[\frac{\ell (\ell+1)}{r^2}+\frac{\left(1-s^2\right)}{r} \left(\frac{2 M}{r^2}-\frac{4 \pi  \left(2 r_s+3 r\right) r_s^4 \rho _s}{3 r^2 \left(r_s+r\right){}^2}-\frac{8 \pi  \left(2 r_s+3
r\right) r_s^4 \rho _s}{3 r \left(r_s+r\right){}^3}+\frac{4 \pi  r_s^4 \rho _s}{r \left(r_s+r\right){}^2}\right)\rt].
\ea
Here $\ell$ is the multipole number. For $s=0$, we obtain potential for the scalar field and $s=1$ yields potential for the vector field. To obtain QNMs which are stationary solutions of Eq. [\r{sch1}], we write \(\Phi(t,r) = e^{-i\omega t} \Phi(r)\) where $\omega$ represents quasinormal frequency. It yields the time-independent Schr$\ddot{o}$dinger-like equation
\begin{equation}
\frac{\partial^{2} \Phi}{\partial r^{*2}} - \left[  \omega^{2} - V_{eff}(r^{*})\right]\Phi = 0.\label{schrodinger}
\end{equation}
We impose the boundary conditions that the waves are purely outgoing at spatial infinity, i.e., $\Phi \sim e^{i\omega r_{\ast}}$ and purely ingoing at the event horizon, i.e., $\Phi \sim e^{-i\omega r_{\ast}}$. It implies that QNMs can only be detected when the source of perturbation is no longer present.\\
The WKB approximation method is widely used to calculate QNMs. It was first presented in \c{Iyer} and was later extended to higher order in \c{Konoplya}. The WKB method is effective for low overtone numbers $n$, especially for $n<\ell$, which is the case in this paper. The $6th$ order WKB formula for QNMs is \c{Konoplya}
\be
\frac{\text{i}(\omega ^{2}-V_{0})}{\sqrt{-2V_{0}^{''}}}-\sum ^{6}_{\text{i}=2}\Omega_\text{i}=n+\frac{1}{2},
\label{WKB}
\ee
where $V_{0}$ is the maximum value of the \textit{Regge--Wheeler} potential obtained at $r_{*}=r_0$, $V''_{0}=\f{d^2 V(r_*)}{dr_{*}^{2}}|_{r_*=r_0}$, and $\Omega_\text{i}$ are WKB corrections \cite{Konoplya}. We display variation of GW frequency and decay rate with core density $\rs$, core radius $r_s$, and multipole number $\ell$ for $n=0$ in Fig. [\r{rea}, \r{ima}, \r{reb}, \r{imb}]. We observe from Fig. [\r{rea}, \r{reb}] that the GW frequency decreases with $\rs$, $\ell$, and $r_s$ for both the perturbations. The nature of variation of decay rate with multipole number for scalar and electromagnetic perturbations, however, are opposite to each other. The decay rate for scalar perturbations decreases with $\ell$ but increases for electromagnetic perturbations. Thus, GWs with higher multipole numbers emitted due to scalar perturbation will propagate further than those caused by electromagnetic perturbation. Fig. [\r{ima}] and [\ref{imb}] show that the decay rate decreases with increasing core redaius or density.
\begin{figure}[H]
\centering
\subfigure[]{
\label{rea1}
\includegraphics[width=0.4\columnwidth]{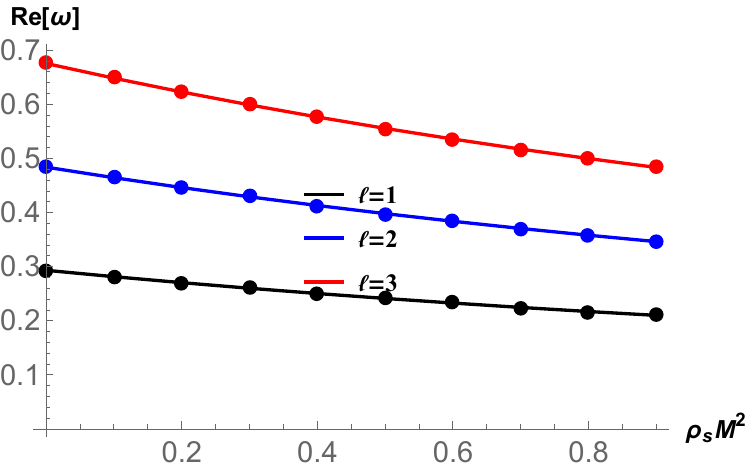}
}
\subfigure[]{
\label{rea2}
\includegraphics[width=0.4\columnwidth]{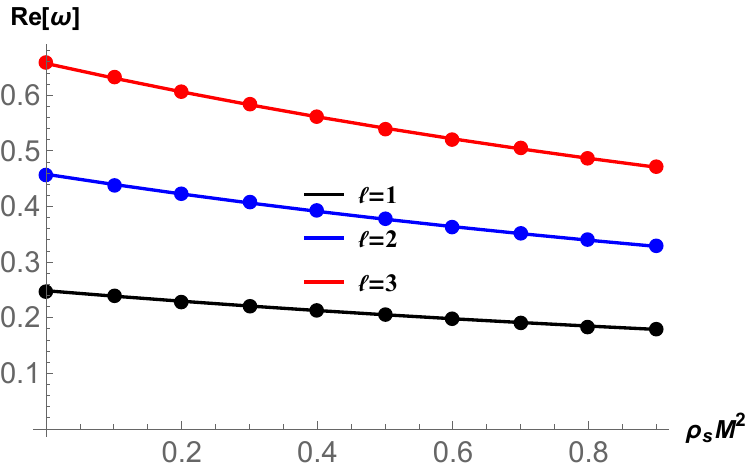}
}
\caption{Variations of gravitational wave frequency with DM density for different $\ell$ keeping $r_s=0.5M$. The left one is for scalar perturbation, and the right one is for electromagnetic perturbation.}
\label{rea}
\end{figure}
\begin{figure}[H]
\centering
\subfigure[]{
\label{ima1}
\includegraphics[width=0.4\columnwidth]{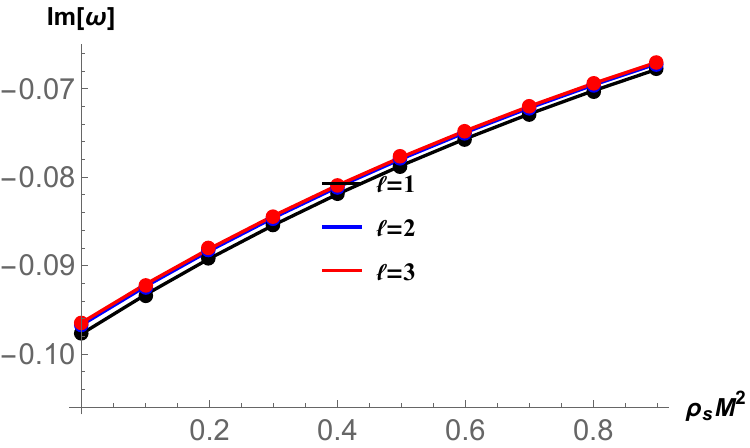}
}
\subfigure[]{
\label{ima2}
\includegraphics[width=0.4\columnwidth]{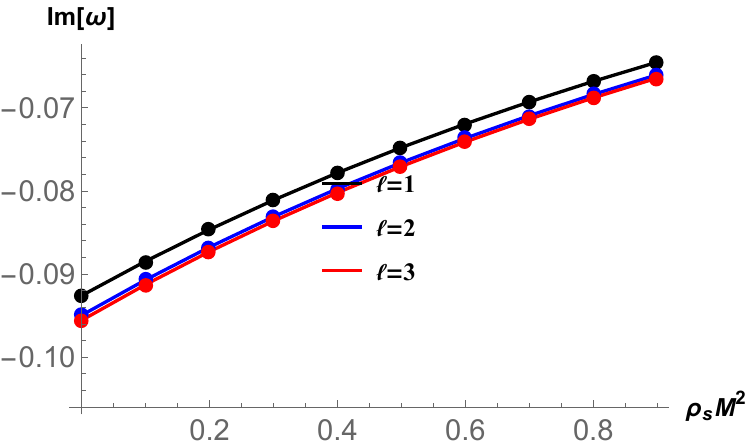}
}
\caption{Variation of decay rate with DM density for different $\ell$ keeping $r_s=0.5M$. The left one is for scalar perturbation, and the right one is for electromagnetic perturbation.}
\label{ima}
\end{figure}
\begin{figure}[H]
\centering
\subfigure[]{
\label{reb1}
\includegraphics[width=0.4\columnwidth]{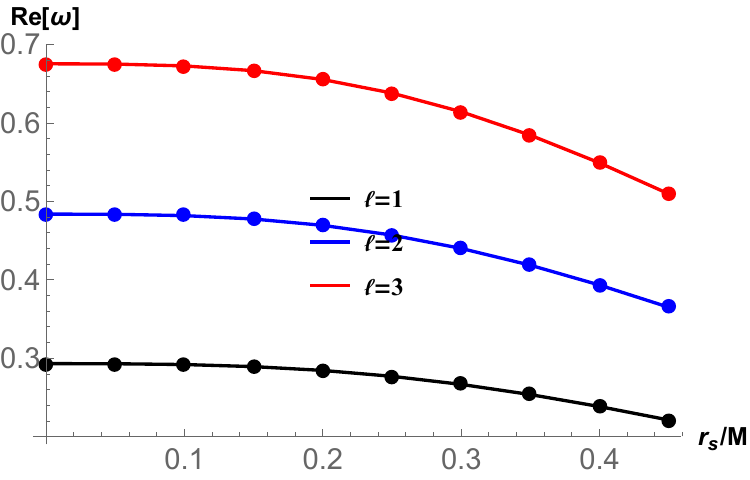}
}
\subfigure[]{
\label{reb2}
\includegraphics[width=0.4\columnwidth]{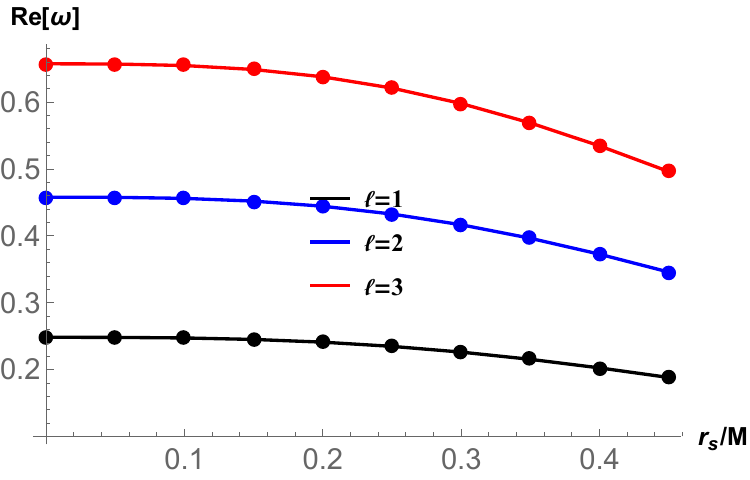}
}
\caption{Variation of gravitational wave frequency with core radius for different $\ell$ keeping $\rs=1.0/M^2$. The left one is for scalar perturbation, and the right one is for electromagnetic perturbation.}
\label{reb}
\end{figure}
We further tabulate numerical values of QNMs for scalar perturbations in Table III for various values of core density with $r_s=0.5M$ and in Table V for various values of core radius with $\rs M^{2}=1.0$. Tables IV and VI provide QNMs for electromagnetic perturbation for various values of $\rs$ and $r_s$ with the same fixed values of $r_s$ and $\rs$, respectively. The negative values of the imaginary part show that the BH system is stable against perturbation. Since the frequency and decay rate decrease with $r_s$ and $\rs$ for both perturbations, GWs emitted from BHs surrounded by DM halo will have lower frequency and decay rate than those emitted from BHs in vaccum. Thus, due to presence of DM halo, GWs emitted will propagate further.
\begin{center}
\begin{tabular}{|c|c|c|c|} 
\hline
$ \rho _s M^2$  & \text{$\ell $=1} & \text{$\ell $=2} & \text{$\ell $=3} \\
\hline
 0. & 0.29291\, -0.0977616 i & 0.483642\, -0.0967661 i & 0.675366\, -0.0965006 i \\
\hline
 0.2 & 0.270069\, -0.0892417 i & 0.445897\, -0.0883635 i & 0.622649\, -0.0881265 i \\
\hline
 0.4 & 0.250013\, -0.0819618 i & 0.412759\, -0.0811762 i & 0.576367\, -0.0809622 i \\
\hline
 0.6 & 0.232342\, -0.0756966 i & 0.383567\, -0.0749851 i & 0.535597\, -0.07479 i \\
\hline
 0.8 & 0.216714\, -0.0702654 i & 0.357752\, -0.0696144 i & 0.499545\, -0.069435 i \\
\hline
 1.0 & 0.202837\, -0.0655236 i & 0.334832\, -0.0649229 i & 0.467538\, -0.0647568 i \\
\hline
\end{tabular}
\captionof{table}{The QNM frequencies of scalar perturbation with several values of $\rs$.} \label{I}
\end{center}

\begin{center}
\begin{tabular}{|c|c|c|c|} 
\hline
$ \rho _s M^2$  & \text{$\ell $=1} & \text{$\ell $=2} & \text{$\ell $=3} \\
\hline
 0. & 0.248191\, -0.092637 i & 0.457593\, -0.0950112 i & 0.656898\, -0.0956171 i \\
\hline
 0.2 & 0.229274\, -0.0846857 i & 0.42213\, -0.086799 i & 0.605798\, -0.0873385 i \\
\hline
 0.4 & 0.21256\, -0.0778638 i & 0.390937\, -0.0797653 i & 0.560895\, -0.0802512 i \\
\hline
 0.6 & 0.19776\, -0.0719728 i & 0.363415\, -0.0737001 i & 0.521308\, -0.0741422 i \\
\hline
 0.8 & 0.184617\, -0.0668517 i & 0.339047\, -0.068434 i & 0.486282\, -0.0688398 i \\
\hline
 1. & 0.172907\, -0.0623703 i & 0.317389\, -0.0638307 i & 0.455168\, -0.0642059 i \\
\hline
\end{tabular}
\captionof{table}{The QNM frequencies of electromagnetic perturbation with several values of $\rs$.} \label{II}
\end{center}

\begin{center}
\begin{tabular}{|c|c|c|c|} 
\hline
$r_s\text{/M }$ & \text{$\ell $=1} & \text{$\ell $=2} & \text{$\ell $=3} \\
\hline
 0. & 0.29291\, -0.0977616 i & 0.483642\, -0.0967661 i & 0.675366\, -0.0965006 i \\
\hline
 0.1 & 0.291747\, -0.0973606 i & 0.481721\, -0.0963695 i & 0.672683\, -0.0961052 i \\
\hline
 0.2 & 0.284244\, -0.0946967 i & 0.469324\, -0.0937378 i & 0.65537\, -0.0934816 i \\
\hline
 0.3 & 0.266475\, -0.0882695 i & 0.439961\, -0.0873923 i & 0.614359\, -0.0871562 i \\
\hline
 0.4 & 0.238217\, -0.0780564 i & 0.393273\, -0.0773077 i & 0.549153\, -0.0771038 i \\
\hline
 0.5 & 0.202837\, -0.0655236 i & 0.334832\, -0.0649229 i & 0.467538\, -0.0647568 i \\
\hline
\end{tabular}
\captionof{table}{The QNM frequencies of scalar perturbation with several values of $r_s$.} \label{III}
\end{center}

\begin{center}
\begin{tabular}{|c|c|c|c|} 
\hline
$r_s\text{/M }$ & \text{$\ell $=1} & \text{$\ell $=2} & \text{$\ell $=3} \\
\hline
 0. & 0.248191\, -0.092637 i & 0.457593\, -0.0950112 i & 0.656898\, -0.0956171 i \\
\hline
 0.1 & 0.247212\, -0.0922588 i & 0.455779\, -0.0946223 i & 0.654291\, -0.0952255 i \\
\hline
 0.2 & 0.24093\, -0.0897568 i & 0.444093\, -0.0920447 i & 0.637481\, -0.0926291 i \\
\hline
 0.3 & 0.226109\, -0.0837342 i & 0.416446\, -0.0858343 i & 0.597687\, -0.0863715 i \\
\hline
 0.4 & 0.20254\, -0.0741586 i & 0.372486\, -0.0759638 i & 0.534414\, -0.0764264 i \\
\hline
 0.5 & 0.172907\, -0.0623703 i & 0.317389\, -0.0638307 i & 0.455168\, -0.0642059 i \\
\hline
\end{tabular}
\captionof{table}{The QNM frequencies of scalar perturbation with several values of $r_s$.} \label{I}
\end{center}
It is evident from our analysis that the signature of Dehnen-type DM can be gauged from the observation of QNMs. We can also distinguish QNMs for scalar and electromagnetic perturbations based on observation. Additionally, we would like to emphasize the efficacy of using the $6th$ order WKB method. Fig. [\r{order}] shows that QNMs converge quickly for pair $n=0,\ell=1$ but fluctuate for $n=3,\ell=0$.
\begin{figure}[H]
\centering
\subfigure[]{
\label{imb1}
\includegraphics[width=0.4\columnwidth]{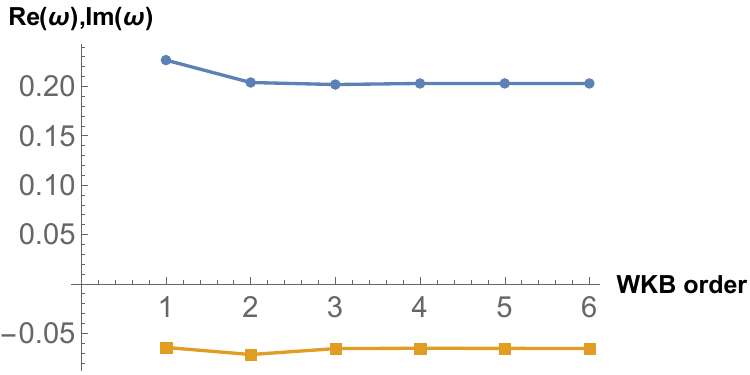}
}
\subfigure[]{
\label{imb2}
\includegraphics[width=0.4\columnwidth]{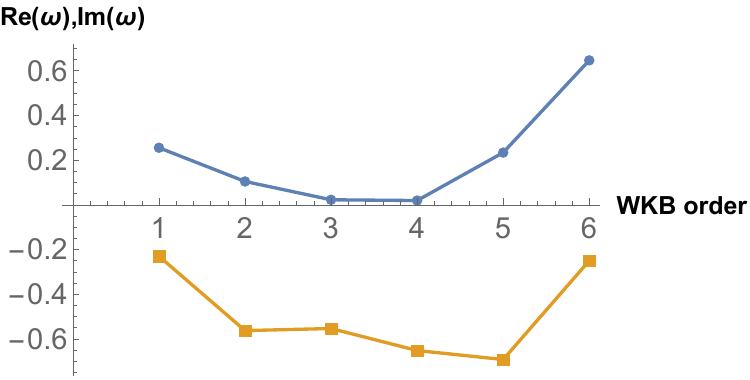}
}
\caption{Variations of the real and imaginary part of quasinormal frequencies with respect to WKB order for various values of $(n,\ell)$ pair. The left one is for (0,1), and the right one is for (3,0). The upper line in each plot is for the real part of quasinormal
modes, and the lower line is for the imaginary part of quasinormal modes. We have taken $\rs=1.0/M^2$ and $r_s=0.5M$.}
\label{order}
\end{figure}
\section{Greybody factor and Hawking spectrum}
Radiation emitted by a BH \c{HAWKING, BEK, KEIF} as observed locally differs from that observed at spatial infinity due to the redshift factor. Greybody distribution provides the Hawking radiation received by an asymptotic observer. The transmission probability of emitted radiation is given by GF. Several different methods can be employed to obtain GF \c{qn32,qn36,qn37,qn38,qn39,qn40,qn41,qn42,qn43,qn44,qn45,qn46}. We will be using the semi-analytic bounds method to obtain lower bounds on GF as \c{GB, GB1, GB2}
\be
T\left( \omega\right) \geq \sec h^{2}\left( \frac{1}{2\omega}\int_{-\infty}^{+\infty }V_{eff}(r_{\ast})dr_{\ast }\right)\Rightarrow T\left( \omega\right) \geq \sec h^{2}\left( \frac{1}{2\omega}\int_{r_{eh}}^{+\infty }V_{eff}(r)\f{dr}{\fr}\right).
\ee
In Fig. [\r{greya}], we exhibit variation of GF with frequency $\omega$ for various values of core density, and Fig. [\r{greyb}] illustrates the dependence of GF on core radius. As evident from these figures, the presence of DM favourably impacts the transmission probability of Hawking radiation. GF asymptotically approaches unity, which is the value for a blackbody.
\begin{figure}[H]
\centering \subfigure[]{ \label{greykr1}
\includegraphics[width=0.45\columnwidth]{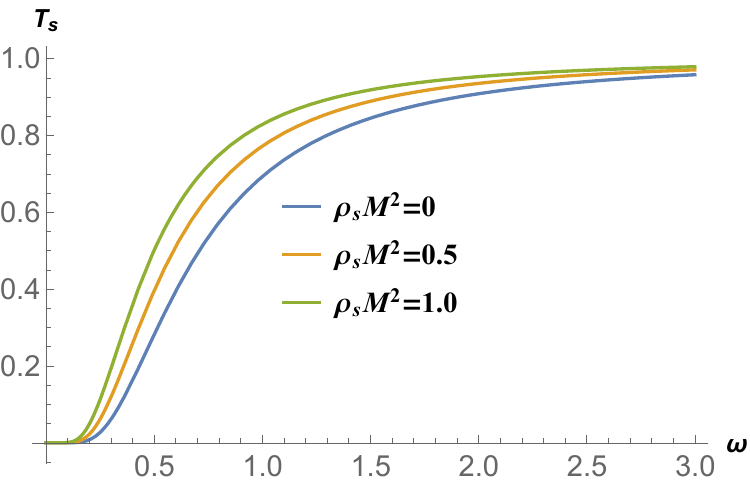}
} \subfigure[]{ \label{greykr2}
\includegraphics[width=0.45\columnwidth]{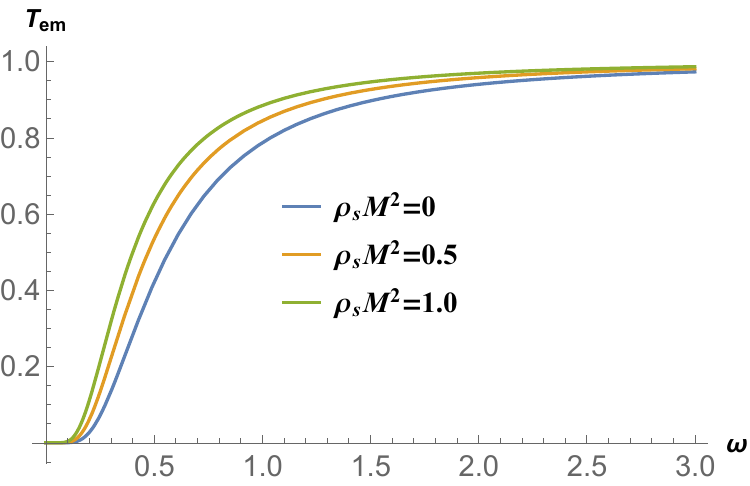}
}
\caption{Greybody bounds for various values of core density. The left one is for the scalar field, and the right one is for the electromagnetic field. We have taken $\ell=1$.}
\label{greya}
\end{figure}

\begin{figure}[H]
\centering \subfigure[]{ \label{greykr1}
\includegraphics[width=0.45\columnwidth]{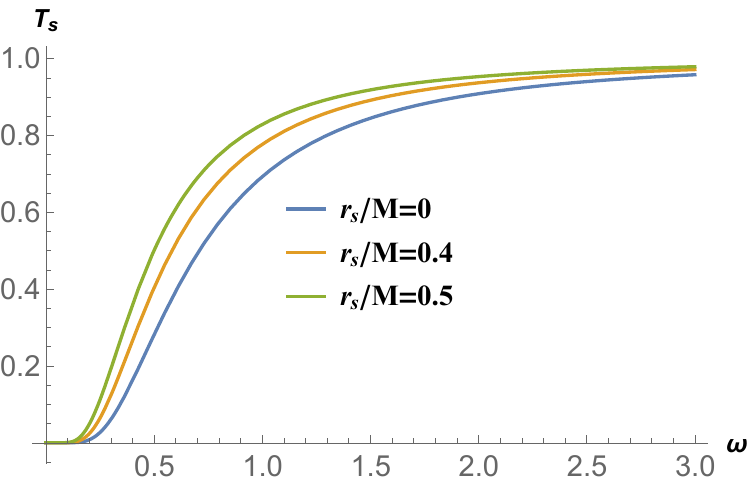}
} \subfigure[]{ \label{greykr2}
\includegraphics[width=0.45\columnwidth]{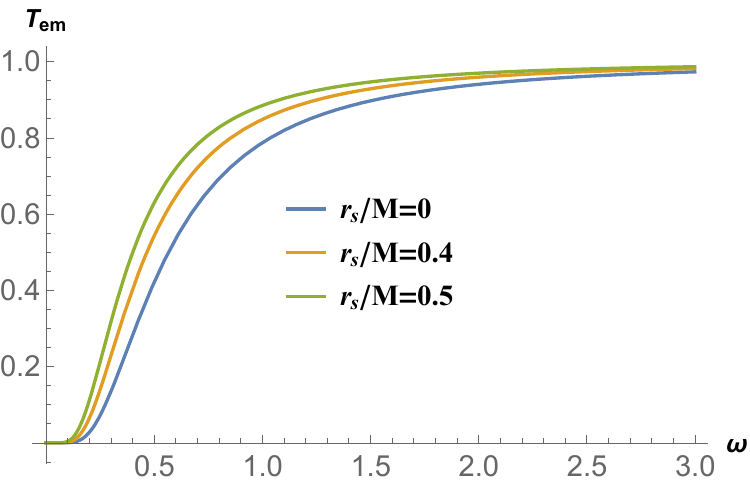}
}
\caption{Greybody bounds for BM BH. The left one is for the scalar field, and the right one is for the electromagnetic field. We have taken $\ell=1$.}
\label{greyb}
\end{figure}
The power emitted by a BH in thermal equilibrium with its surrounding in the $\ell th$ mode is \c{yg2017, fg2016}
\begin{equation}\label{pl}
P_\ell\left(\omega\right)=\frac{A}{8\pi^2}T(\omega)\frac{\omega^3}{e^{\omega/T_{H}}-1}.
\end{equation}
Here, $A$ is taken to be the horizon area \cite{yg2017}, and $T_{H}$ is the Hawking temperature \c{kalyan}.
\begin{figure}[H]
\centering \subfigure[]{ \label{greykr1}
\includegraphics[width=0.45\columnwidth]{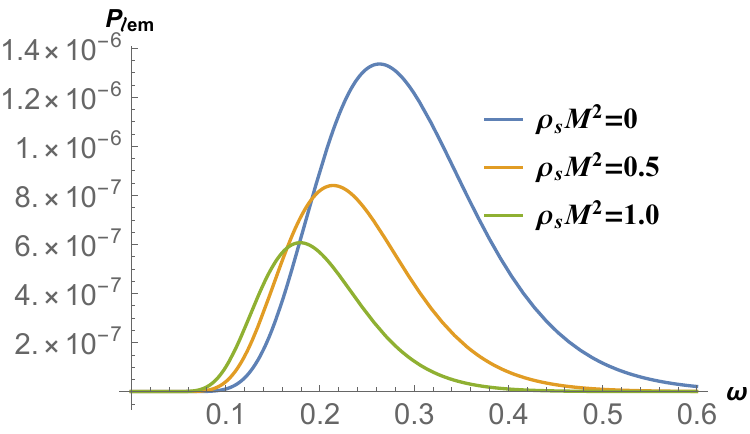}
} \subfigure[]{ \label{greykr2}
\includegraphics[width=0.45\columnwidth]{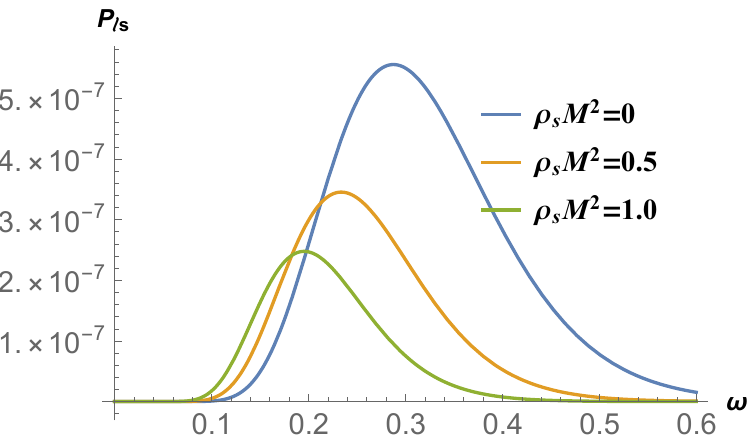}
}
\caption{Variation of power emitted in the $\ell$th mode with $\omega$ for different values of $\rs$ keeping $r_s$ fixed at $0.5M$. The left one is for the electromagnetic field, and the right one is for the scalar field. We have taken $\ell=1$.}
\label{pla}
\end{figure}

\begin{figure}[H]
\centering \subfigure[]{ \label{greykr1}
\includegraphics[width=0.45\columnwidth]{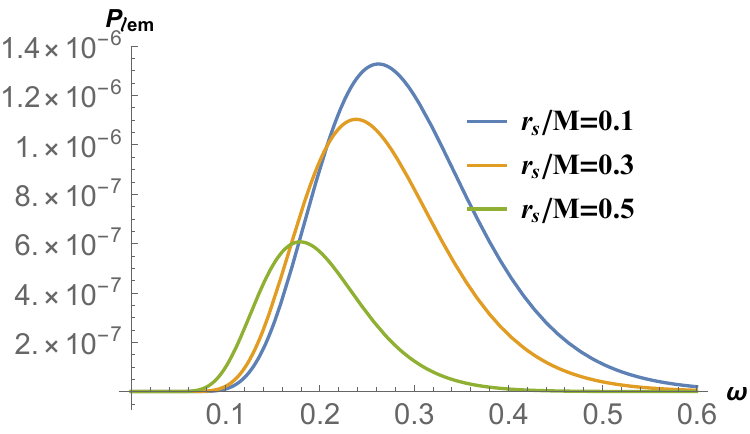}
} \subfigure[]{ \label{greykr2}
\includegraphics[width=0.45\columnwidth]{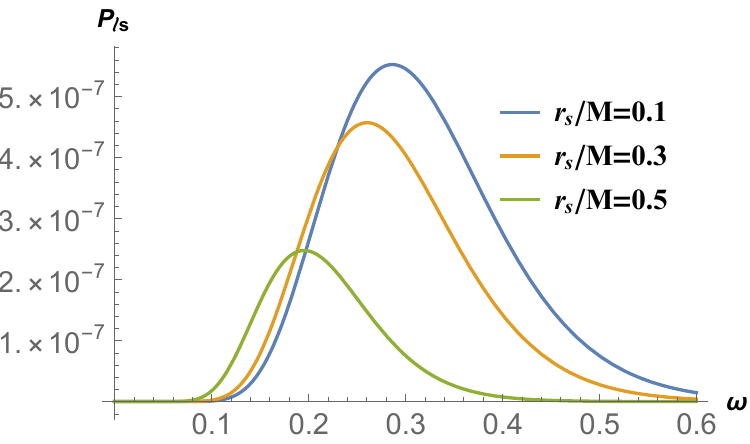}
}
\caption{Variation of power emitted in the $\ell$th mode with $\omega$ for different values of $r_s$ keeping $\rs$ fixed at $1.0/M^2$. The left one is for the electromagnetic field, and the right one is for the scalar field. We have taken $\ell=1$.}
\label{plb}
\end{figure}
Fig. [\r{pla}] illustrates the impact of core density on the power emitted, and Fig. [\r{plb}] shows the dependence of power emitted on core radius. It exhibits the adverse impact of DM on thermal radiation. The power emitted decreases, and the peak shifts towards the left with an increase in core density or radius for both the perturbations. This is due to the fact that the Hawking temperature decreases with core density and radius. We can, therefore, conclude that thermal radiation received from a \s in the vacuum will be greater than that received from a \s embedded in a Dehnen-type DM halo.
\section{weak gravitational lensing}
\begin{figure}[H]
\begin{center}
\includegraphics[scale=0.7]{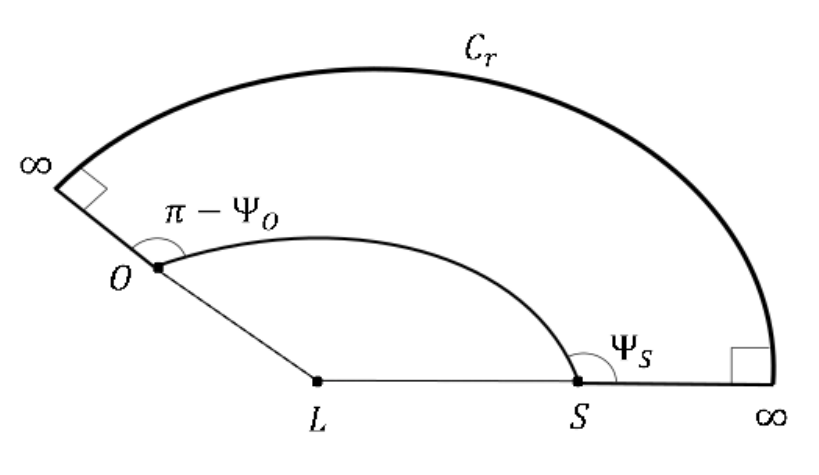}
\caption{Schematic diagram of the quadilateral ${}_O^{\infty}\Box_{S}^{\infty}$. }\label{lensing}
\end{center}
\end{figure}

The motivation behind studying gravitational lensing is the dependence of lensing angle on the inherent characteristics of the underlying spacetime. It is for this reason that gravitational lensing is such a widely researched topic. We intend to analyze the effect of DM on lensing. We will be using the Gauss-Bonnet theorem to study weak gravitational lensing \c{GW}. The formula for the gravitational lensing is \c{ISHIHARA1, CARMO}
\be
\gamma_D=-\int\int_{{}_R^{\infty}\Box_{S}^{\infty}} K
dS,\label{deflectionangle}
\ee
where ${}_O^{\infty}\Box_{S}^{\infty}$ is the quadrilateral (please see Fig. (\ref{lensing})) and K is the Gaussian curvature. Considering null geodesics, one obtains the Gaussian curvature as \c{WERNER}
\ba\nn
K&=&-\f{\fr^{3/2}}{r}\f{d}{d\text{r}}\lt(\fr \f{d}{d\text{r}}(\f{r}{\sqrt{\fr}})\rt),\\\nn
&=&\frac{3 M^2}{r^4}-\frac{2 M}{r^3}+\left(\frac{8 \pi  M}{r^4}-\frac{8 \pi }{3 r^3}\right) r_s^3 \rho _s+\mc{O}\left(\f{\rs r_s^4}{r^4}, \f{M \rs r_s^4}{r^5} \rt).
\ea

We rewrite Eq. (\ref{deflectionangle}) as \c{ONO1}
\be
\int\int_{{}_O^{\infty}\Box_{S}^{\infty}} K dS =
\int_{\phi_S}^{\phi_O}\int_{\infty}^{r_0} K \sqrt{\zeta}dr
d\phi,\label{Gaussian}
\ee
where $r_0$ is the least distance from BH. We first consider a straight-line trajectory and obtain the deflection angle. Here, the path is given by $r=\frac{b}{sin\phi}$. Our initial deflection angle using straight-line trajectory is $\gamma_D^0=\frac{4 M}{b}+\frac{3 \pi  M^2}{4 b^2}+\left(\frac{32 \pi  M^2}{3 b^3}+\frac{2 \pi ^2 M}{b^2}+\frac{16 \pi }{3 b}\right) r_s^3 \rho _s+\mc{O}\lt(\frac{M^3}{b^3},\frac{M^3\rs r_s^3}{b^4}\rt)$. To obtain higher order correction terms, we will use the trajectory \c{CRISNEJO} 
\be
u=\f{1}{r}=\frac{\sin\phi}{b} + \frac{M(1-\cos\phi)^2}{b^2}-\frac{M^2(60\phi\,
\cos\phi+3\sin3\phi-5\sin\phi)}{16b^3}+\mathcal{O}\left( \frac{M^2\alpha}{b^5}\right),
,\label{uorbit}
\ee
where $u=1/r$. The deflection for the above trajectory is 
\be
\gamma_D=\frac{4 M}{b}+\frac{15 \pi  M^2}{4 b^2}+\frac{128 M^3}{3 b^3}+\left(\frac{239 \pi ^2 M^3}{2 b^4}+\frac{224 \pi  M^2}{3 b^3}+\frac{6 \pi ^2 M}{b^2}+\frac{16 \pi }{3 b}\right) r_s^3 \rho
   _s+\mc{O}\lt(\frac{M^4}{b^4},\frac{M^4\rs r_s^3}{b^5}\rt),\label{angle}
\ee
where we have integrated $\phi$ from $0$ to $\pi+\gamma_D^0$. Fig. [\r{def}] illustrates the effect of DM parameters on the deflection angle. 
\begin{figure}[H]
\centering \subfigure[]{ \label{greykr1}
\includegraphics[width=0.45\columnwidth]{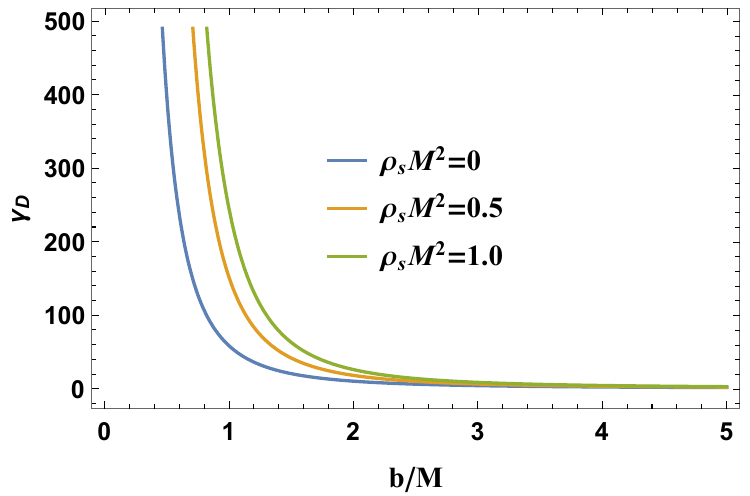}
} \subfigure[]{ \label{greykr2}
\includegraphics[width=0.45\columnwidth]{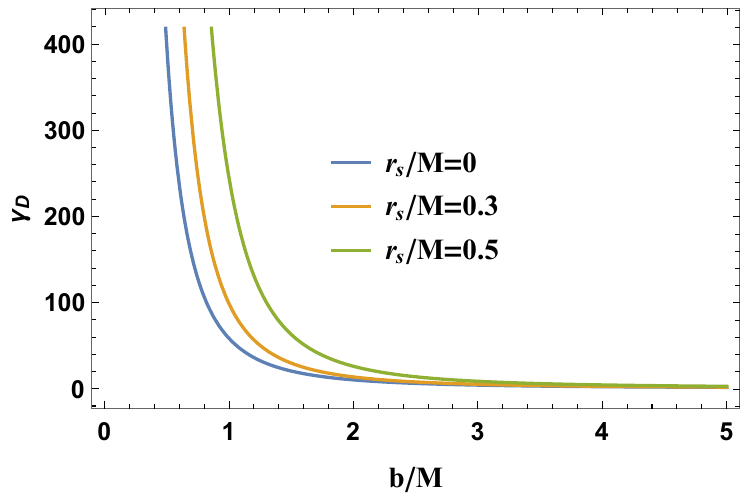}
}
\caption{Deflection angle for weak gravitational lensing. The left
panel is for different values of core density keeping $r_s=0.5M$ and the right one for different values of core radius keeping $\rs=1.0/M^2$ .}\label{def}
\end{figure}
The angle of deflection decreases with the impact parameter as the gravitational pull of the central massive object reduces with distance away from it. From expression [\ref{angle}] we observe that the deflection angle increases linearly with core density for a fixed value of core radius, whereas, the dependence of the angle on the core radius is cubic order where core density is kept fixed. Our analysis in this section makes it clear that we can gauge the impact of Dehnen type DM from lensing observation as the impact of DM on the angle of deflection is significant. 
\section{parameter estimation using shadow observable}
This section will utilize recent images and corresponding observables related to SMBHs $M87^*$ and $Sgr A^*$ to constrain two parameters - core radius $r_s$ and core density $\rs$. Constraints regarding deviation from \s, $\delta$, for $M87^*$ obtained from EHT \c{M871, M872} and $Sgr A^*$ obtained from Keck and VLTI observatories \c{keck, vlti1, vlti2} will be used here. We first define the observable $\delta$ as \c{del}
\be
\delta=\f{R_s}{3\sqrt{3}M}-1.
\ee
$R_s$ is the shadow radius for the BH under consideration. $3\sqrt{3}M$ is the shadow radius for \s BH. A positive value of $\delta$ will indicate a larger shadow radius for our BH than \s, whereas negative values will infer a smaller shadow. Bounds on deviation parameter $\delta$ obtained for $M87^*$ and $Sgr A^*$ are tabulated in Table [\r{bounds}]. Our metric will be put to test using these bounds.
\begin{center}
\begin{tabular}{|l|c|c|c|r|}
\hline
BH & Observatory & $\delta$ & 1$\sigma$ bounds & 2$\sigma$ bounds\\
\hline
\hline
$M87^*$ & EHT & $-0.01^{+0.17}_{-0.17}$ & $4.26\le \frac{R_s}{M}\le 6.03$ &  $3.38\le \frac{R_s}{M}\le 6.91$\\
\hline
\hline
\multirow{2}{*}{$Sgr A^*$}&{VLTI} & $-0.08^{+0.09}_{-0.09}$ &{$4.31\le \frac{R_s}{M}\le 5.25$} &{$3.85\le \frac{R_s}{M}\le 5.72$}\\[3mm]
& Keck & $-0.04^{+0.09}_{-0.10}$ & {$4.47\le \frac{R_s}{M}\le 5.46$} & {$3.95\le \frac{R_s}{M}\le 5.92$}\\[1mm]
\hline
\end{tabular}
\captionof{table}{Bounds on $\delta$ from different observatories.} \label{bounds}
\end{center}
We display variations of deviation parameter with core density in Fig. [\r{del1}] and with core radius in Fig. [\r{del2}] along with upper $1\sigma$ and $2\sigma$ bounds obtained from EHT, Keck, and VLTI observations. $\delta$ increases almost linearly with increasing $\rs$, whereas, the change in $\delta$ with $r_s$ is much steeper. 

\begin{figure}[H]
\centering \subfigure[]{ \label{del1}
\includegraphics[width=0.45\columnwidth]{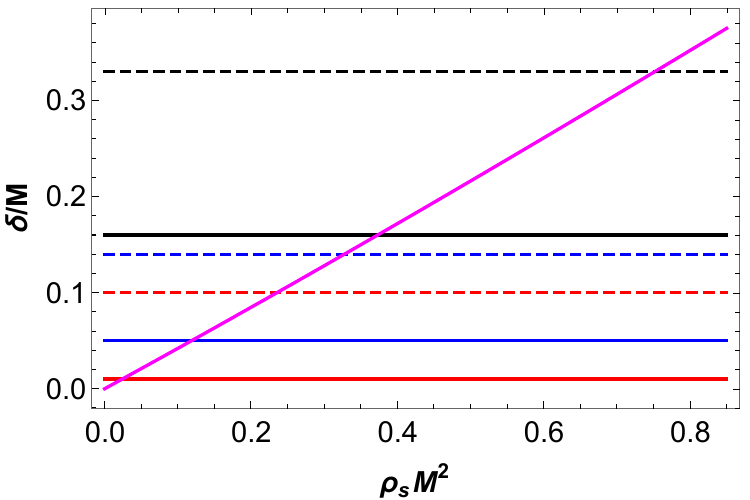}
} \subfigure[]{ \label{del2}
\includegraphics[width=0.45\columnwidth]{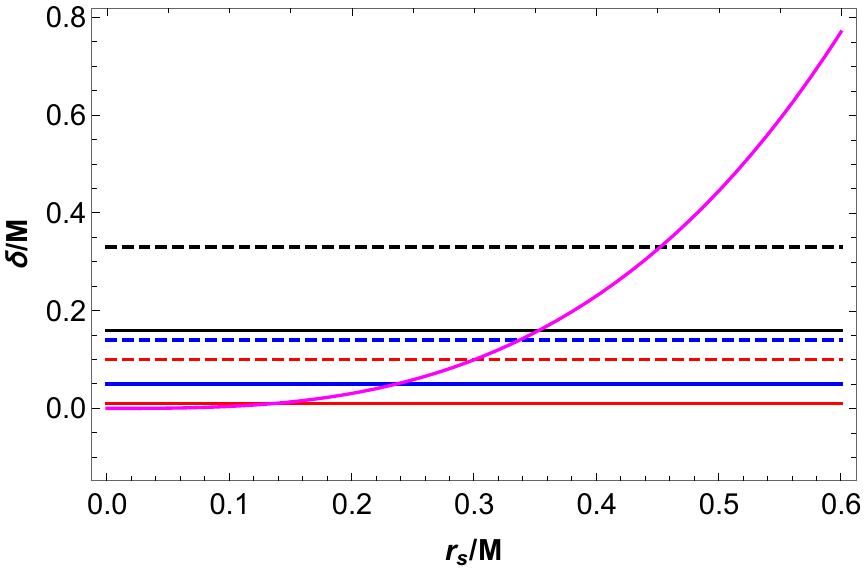}
}
\caption{Variations of deviation parameter shown with respect to core density keeping $r_s=0.5M$ in the left panel and with respect to core radius keeping $\rs M^2=1.0$ in the right panel. Horizontal solid lines are 1-$\sigma$ upper bound, and dashed lines are 2-$\sigma$ upper bounds. The color code for horizontal lines is: Black is for EHT, Blue is for Keck, and Red is for VLTI. }\label{del}
\end{figure}
Horizontal lines in Fig. [\r{del}] provide upper $1\sigma$ and $2\sigma$ bounds on $\delta$. Their intersection with variation curve indicate existence of upper bounds on $\rs$ and $r_s$. We have obtained upper bounds on core radius and core density for different observations. Our study shows that keeping $\rs M^2=1.0$, $r_s\le 0.353377M$ within $1\sigma$ confidence level and $r_s \le 0.452259M$ within $2\sigma$ for $M87^*$. Similarly, upper bounds on core radius consistent with Keck and VLTI observations are $0.236933M$ and $0.13662M$ within $1 \sigma$ confidence level and $0.337556M$ and $0.300702M$ within $2 \sigma$ confidence level. Upper bounds on core density are obtained keeping core radius fixed at $0.5M$. Upper bounds on $\rs$ are as follows:\\
From EHT observation $\rs M^2 \le 0.373203$ within $1 \sigma$ and $\rs M^2 \le 0.795563$ within $2 \sigma$.\\
From Keck observation, $\rs M^2 \le 0.118825$ within $1 \sigma$ bounds and $\rs M^2 \le 0.327586$ within $2 \sigma$ bounds. \\
VLTI observation yields $\rs M^2 \le 0.0239478$ within $1 \sigma$ and $\rs M^2 \le 0.235545$ within $2 \sigma$.\\
 These results are tabulate in [\r{bounds1}].
\begin{center}
\begin{tabular}{|l|c|c|c|c|r|} 
\hline
\multicolumn{2}{|c}{} &\multicolumn{2}{c}{bounds on $r_s/M$} & \multicolumn{2}{|c|}{bounds on $\rs M^2$}\\
\hline
BH & Observatory &  1$\sigma$ bounds & 2$\sigma$ bounds &  1$\sigma$ bounds & 2$\sigma$ bounds\\
\hline
\hline
$M87^*$ & EHT & $0.353377$ & $0.452259$ & $ 0.373203$  & $0.795563$ \\
\hline
		\hline
\multirow{2}{*}{$Sgr A^*$}&{VLTI} & $0.13662$ & $0.300702$ & $ 0.0239478$ & $ 0.235545$ \\[3mm]
         & Keck & $0.236933$ & $0.337556$ & $ 0.118825$ & $ 0.327586$\\[1mm]
\hline
\end{tabular}
\captionof{table}{Bounds on $r_s/M$ and $\rs M^2$ from different observatories.} \label{bounds1}
\end{center}

Table [\r{bounds1}] shows the upper bounds of core radius and density for fixed values of core density and radius. To have a better idea of the viability of our model, we exhibit variation of deviation parameter $\delta$ with core radius and density for $M87^*$ and $Sgr A^*$.

\begin{figure}[H]
\begin{center}
\begin{tabular}{cccc}
\includegraphics[width=0.4\columnwidth]{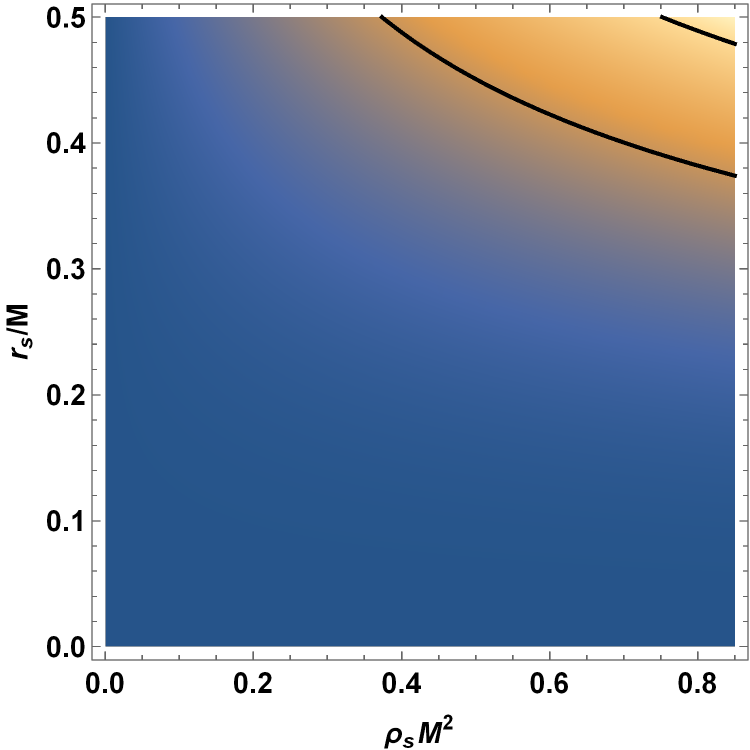}&
\includegraphics[width=0.05\columnwidth]{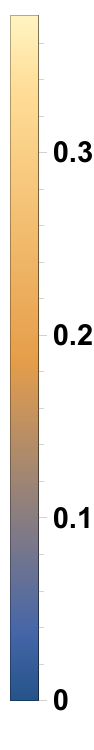}&
\includegraphics[width=0.4\columnwidth]{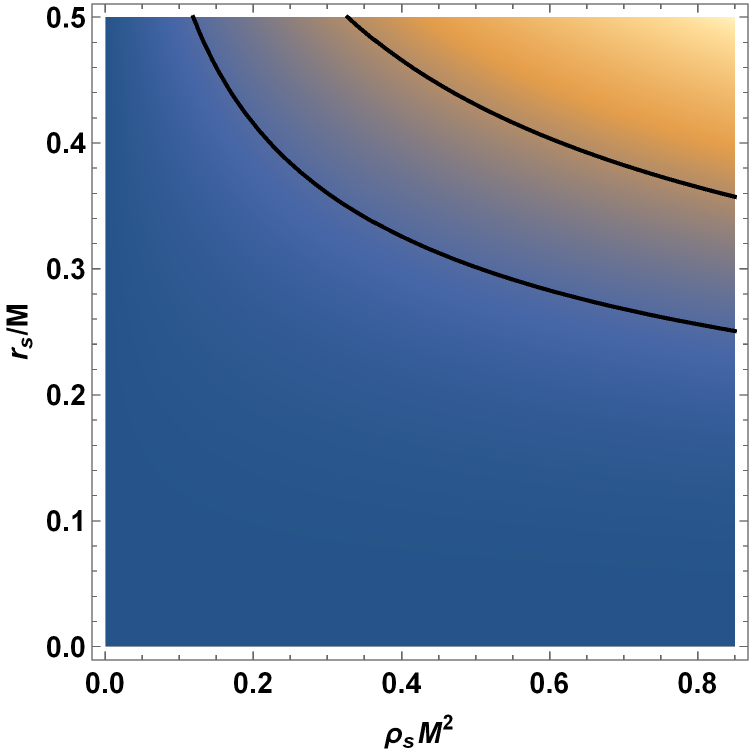}&
\includegraphics[width=0.05\columnwidth]{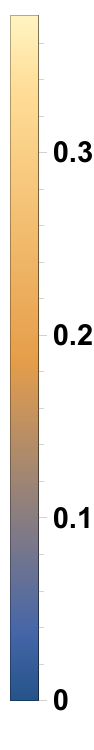}
\end{tabular}
\caption{Variation of deviation parameter $\delta$ with core $\rs$ and $r_s$. The left one is for $M87^*$ with EHT constraints, and the right one is for $Sgr A^*$ with Keck bounds. In each plot, the upper solid black line corresponds to the upper $2\sigma$ bound, and the lower one is for the upper $1 \sigma$ bound. }\label{para1}
\end{center}
\end{figure}

\begin{figure}[H]
\begin{center}
\begin{tabular}{cc}
\includegraphics[width=0.4\columnwidth]{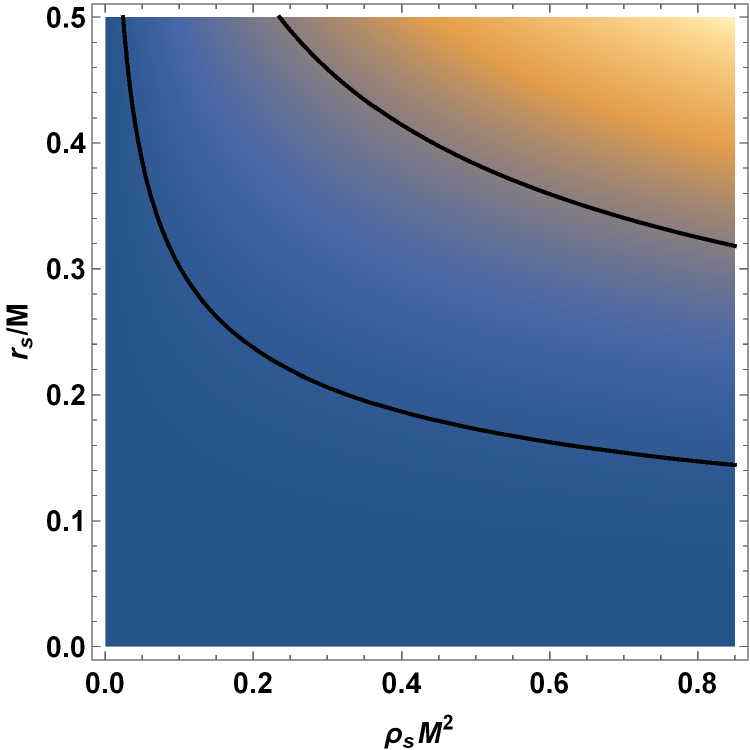}&
\includegraphics[width=0.05\columnwidth]{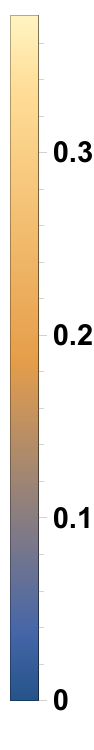}
\end{tabular}
\caption{Variation of deviation parameter $\delta$ with core $\rs$ and $r_s$ for $Sgr A^*$ with VLTI bounds. The upper solid black line corresponds to the upper $2\sigma$ bound, and the lower one is for the upper $1 \sigma$ bound. }\label{para2}
\end{center}
\end{figure}
Figs. [\r{para1}, \r{para2}] show that for finite parameter space $(\rs M^2 - r_s/M)$, our BH under consideration is congruent with experimental data from EHT, Keck, and VLTI observatories. We can, therefore, conclude that a \s BH embedded in a Dehnen-type DM halo is a viable candidate for SMBH.
\section{conclusions}
We have extensively studied the impact of Dehnen-type DM on null and time-like geodesics, QNMs, greybody factor, Hawking temperature, and weak GL. We have also tried to constrain DM parameters, namely, core density and radius, using $M87^*$ and $Sgr A^*$ data. Utilizing Lagrangian formulation, we first obtain the potential for the motion of mass-less and massive particles in the background of our BH. Then, imposing appropriate conditions on the potential, we obtain equations whose solutions yield photon radius and ISCO radius. Photon radius, along with imposed conditions on the potential, provides a critical impact parameter, which is also the shadow radius for an asymptotic observer.  We exhibit the qualitative and quantitative nature of variation of observables with respect to DM parameters. We observe that a \s BH embedded in a DDM halo cast a larger shadow than a \s BH in the vacuum. The ISCO radius, too, is found to be larger for SDDM. Even though the extent of the impact of core radius and density on these observables are different, their nature of effect is the same, i.e., increasing core radius or density increases photon and ISCO radii.\\

Next, QNMs, with the help of the $6th$ order WKB method, are studied. Since we have confined ourselves to the case of $n=0$, the $6th$ order WKB method provides accurate results. We have considered scalar and electromagnetic perturbations. The frequency of GWs emitted is found to be decreasing with DM parameters for both the perturbations. The imaginary part of QNMs is always negative, indicating the stability of the BH-matter combination against perturbation. Although GWs with higher multipole numbers always have higher frequency irrespective of the type of perturbation, the situation is not similar with regard to the decay rate. While for scalar perturbation, emitted GWs decay faster for smaller multipole numbers, the situation is reversed in the case of electromagnetic perturbation. This makes GWs with higher multipole numbers originating from BHs perturbed due to scalar field and GWs with lower multipole numbers originating from BHs perturbed due to electromagnetic field move further and live longer. We can also infer from our study that GWs emitted from BHs embedded in DM halo decay slower and hence can move further. Since the frequency and decay rate of GWs emitted due to scalar and electromagnetic perturbations are significantly different, we can differentiate the two perturbations based on QNM observation.\\

We then move on to study the impact of DM on the greybody factor and Hawking spectrum for scalar and electromagnetic perturbations. The greybody factor provides information regarding the transmission probability of radiation to be observed at spatial infinity. Since the greybody factor increases with core density and radius, the probability of radiation getting transmitted to an asymptotic observer increases. It is evident from Fig. [\r{pla}, \r{plb}] that a reduction in the Hawking radiation received by an asymptotic observer occurs due to an increase in core radius and density for both types of perturbation. This is caused by reduction in the Hawking temperature due to DM. We apprehend the impact of DM parameters on weak GL using the Gauss-Bonnet theorem. We have obtained higher-order correction terms in the deflection angle where terms up to the cubic order in $M$ are retained. The deflection angle gets enhanced due to the presence of DM.\\

Finally, we have used dounds on deviation from \s $\delta$ for $M87^*$ and $Sgr A^*$ to probe the viability of our model and put constrain on DM parameters - core radius and density. The parameter $\delta$ for our BH never reaches its lower bounds since the deviation parameter is always positive, indicating a larger BH shadow. The deviation parameter is an increasing function of core radius and density. We have displayed parameter space $(\rs M^2 - r_s/M)$ for $\delta$ with regard to $M87^*$ and $Sgr A^*$. It clearly shows that for finite parameter space, our model satisfies experimental observations. We have also obtained $1\sigma$ and $2\sigma$ upper bounds on core radius and core density for a fixed value of core density and core radius, respectively. Upper bounds on $\rs$ and $r_s$ are tabulated in [\ref{bounds1}]. These observations make SDDM a feasible candidate for SMBH. We, in our next endeavor, intend to study the rotating counterpart of SDDM and apprehend the DM effect when an additional parameter, i.e., the BH spin comes into play.


\begin{thebibliography}{the}
\bibitem{rubin} V. C. Rubin, J. Ford, W. K., and N. Thonnard, Astrophys. J. 238, 471 (1980).
\bibitem{persic} M. Persic, P. Salucci, and F. Stel, Mon. Not. R. Astron. Soc. 281, 27 (1996), arXiv:astro-ph/9506004 [astro-ph].
\bibitem{akiyamal1} K. Akiyama and et al. (Event Horizon Telescope Collaboration), Astrophys. J. 875, L1 (2019), arXiv:1906.11238 [astro-ph.GA] .
\bibitem{akiyamal6} K. Akiyama and et al. (Event Horizon Telescope Collaboration), Astrophys. J. 875, L6 (2019), arXiv:1906.11243 [astro-ph.GA] .
\bibitem{sofue} Y. Sofue, “Mass Distribution and Rotation Curve in the Galaxy",  in Planets, Stars and Stellar Systems. Volume 5: Galactic Structure and Stellar Populations, Vol. 5, edited by T. D. Oswalt and G. Gilmore (2013) p. 985.
\bibitem{boshkaye} K. Boshkayev and D. Malafarina, Mon. Not. R. Astron. Soc. 484, 3325 (2019), arXiv:1811.04061 [gr-qc] .
\bibitem{kiselev} V. V. Kiselev, arXiv e-prints (2003), arXiv:gr-qc/0303031 [gr-qc] .
\bibitem{li} M.-H. Li and K.-C. Yang, Phys. Rev. D 86, 123015 (2012), arXiv:1204.3178 [astro-ph.CO] .
\bibitem{xu} Z. Xu, X. Hou, and J. Wang, Class. Quantum Grav. 35, 115003 (2018), arXiv:1711.04538 [gr-qc] .
\bibitem{haroon} S. Haroon, M. Jamil, K. Jusufi, K. Lin, and R. B. Mann, Phys. Rev. D 99, 044015 (2019), arXiv:1810.04103 [gr-qc] .
\bibitem{ra} R. A. Konoplya, Phys. Lett. B 795, 1 (2019), arXiv:1905.00064 [gr-qc].
\bibitem{hendi} S. H. Hendi, A. Nemati, K. Lin, and M. Jamil, Eur. Phys. J. C 80, 296 (2020), arXiv:2001.01591 [gr-qc] .
\bibitem{kj} K. Jusufi, M. Jamil, P. Salucci, T. Zhu, and S. Haroon, Phys. Rev. D 100, 044012 (2019).
\bibitem{bz} B. Narzilloev, J. Rayimbaev, S. Shaymatov, A. Abdujabbarov, B. Ahmedov, and C. Bambi, Phys. Rev. D 102, 104062 (2020), arXiv:2011.06148 [gr-qc] .
\bibitem{ss} S. Shaymatov, B. Ahmedov, and M. Jamil, Eur. Phys. J. C 81, 588 (2021).
\bibitem{rayimbaev} J. Rayimbaev, S. Shaymatov, and M. Jamil, Eur. Phys. J. C 81, 699 (2021), arXiv:2107.13436 [gr-qc] .
\bibitem{dehnen} W. Dehnen, A family of potential–density pairs for spherical galaxies and bulges, Mon. Not. R. Astron. Soc. 265, 250 (1993).
\bibitem{mo} H. Mo, F. van den Bosch, and S. White, Galaxy Formation and Evolution (Cambridge University Press, Cambridge, England, UK, 2010).
\bibitem{kalyan} Mrinnoy M. Gohain, Prabwal Phukon, Kalyan Bhuyan, arXiv: 2407.02872
\bibitem{RC} Reggie C. Pantig, Paul K. Yu, Emmanuel T. Rodulfo, and Ali O¨ vgu¨n, “Shadow and weak deflection angle of extended uncertainty principle
black hole surrounded with dark matter,” Annals Phys. 436, 168722 (2022), arXiv:2104.04304 [gr-qc].
\bibitem{RA} R. A. Konoplya and A. Zhidenko, “Solutions of the Einstein Equations for a Black Hole Surrounded by a Galactic Halo,” Astrophys. J.
933, 166 (2022), arXiv:2202.02205 [gr-qc].
\bibitem{RA1} R. A. Konoplya, “Shadow of a black hole surrounded by dark matter,” Phys. Lett. B 795, 1–6 (2019), arXiv:1905.00064 [gr-qc].
\bibitem{ZX} Zhaoyi Xu, Xian Hou, Xiaobo Gong, and Jiancheng Wang, “Black Hole Space-time In Dark Matter Halo,” JCAP 09, 038 (2018).
\bibitem{ZX1} Zhaoyi Xu, Xiaobo Gong, and Shuang-Nan Zhang, “Black hole immersed dark matter halo,” Phys. Rev. D 101, 024029 (2020).
\bibitem{RC1} Reggie C. Pantig and Emmanuel T. Rodulfo, “Rotating dirty black hole and its shadow,” Chin. J. Phys. 68, 236–257 (2020),
arXiv:2003.06829 [gr-qc].
\bibitem{WJ} Wajiha Javed, Hafsa Irshad, Reggie C. Pantig, and Ali vgn, “Weak deflection angle by kalb-ramond traversable wormhole in plasma and
dark matter mediums,” Universe 8 (2022), 10.3390/universe8110599.
\bibitem{WJ1} Wajiha Javed, Sibgha Riaz, Reggie C. Pantig, and Ali O¨ vgu¨n, “Weak gravitational lensing in dark matter and plasma mediums for
wormhole-like static aether solution,” Eur. Phys. J. C 82, 1057 (2022), arXiv:2212.00804 [gr-qc].
\bibitem{KJ} Kimet Jusufi, Mubasher Jamil, and Tao Zhu, “Shadows of Sgr $A^*$ black hole surrounded by superfluid dark matter halo,” Eur. Phys. J. C
80, 354 (2020), arXiv:2005.05299 [gr-qc].
\bibitem{SN} Sourabh Nampalliwar, Saurabh Kumar, Kimet Jusufi, Qiang Wu, Mubasher Jamil, and Paolo Salucci, “Modeling the Sgr A* Black Hole Immersed in a Dark Matter Spike,” Astrophys. J. 916, 116 (2021), arXiv:2103.12439 [astro-ph.HE].
\bibitem{sanjar} Sanjar Shaymatov, Daniele Malafarina, Bobomurat Ahmedov: Physics of the Dark Universe 34, 100891 (2021).
\bibitem{bakhtiyor} Bakhtiyor Narzilloev et al. : Phys. Rev. D 102, 104062 (2020).
\bibitem{panting} Reggie C. Pantig et al. : Annals of Physics 436, 168722 (2022).
\bibitem{sanjar1} Sanjar Shaymatov, Pankaj Sheoran, Sanjay Siwach: Phys. Rev. D 105, 104059 (2022).
\bibitem{nozari} Kourosh Nozari, Sara Saghafi, Fateme Aliyan: Eur. Phys. J. C, 83, 449 (2023).
\bibitem{faraji} Shokoufe Faraji, Joao Luis Rosa: arXiv:2403.02597 [gr-qc].
\bibitem{sobhan} Sobhan Kazempour, Sichun Sun, Chengye Yu: arXiv:2404.11333 [gr-qc].
\bibitem{KONOR} R. A. Konoplya and A. Zhidenko, Rev. Mod. Phys. 83, 793 (2011)
\bibitem{13}J. S. F. Chan and R. B. Mann, Phys. Rev. D 55  7546
(1997)
\bibitem{14} G.T. Horowitz , V. E. Hubeny, Phys. Rev. D62
024027  (2000
\bibitem{15} S. Hod, Phys. Rev. Lett. 81  4293 (1998)
\bibitem{16} R. A. Konoplya and A. Zhidenko, Rev. Mod. Phys. 83
793  (2011)
\bibitem{17} B. Chen and J. Zhang, Phys. Rev. 84  124039 (2011)
arXiv:1110.3991 [hep-th]
\bibitem{18} Y. Kim, Y. S. Myung and Y. Park, Eur. Phys. J. C 73 138 (2013)
\bibitem{19} I. Z. Stefanov, S. S. Yazadjiev and G. G. Gyulchev,  Phys. Rev. Lett. 104
251103 (2010)
\bibitem{20} R. A. Konoplya, A. F. Zinhailo,  Z. StuchlikPhys. Rev. D 102, 044023 (2020)
\bibitem{KONO} R. A. Konoplya and A. Zhidenko, Phys. Rev. D 105, 104032 (2022)
\bibitem{KONO1} K. A. Bronnikov, R. A. Konoplya, and T. D. Pappas, Phys. Rev. D 103, 124062 (2021)
\bibitem{KONO2} R. A. Konoplya, Phys. Rev. D 103, 044033 (2021).
\bibitem{21} T. V. Fernandes, D. Hilditch, J. P. S. Lemos,  V. Cardoso, Phys. Rev. D 105 044017 (2022)
\bibitem{22} K. Jusu, M. Azreg-Anou, M. Jamil, Shao-Wen Wei, Q. Wu,  A. Wang, Phys. Rev. D 103, 024013 (2021)
 \bibitem{23} E. Franzin, S. Liberati, J. Mazza, R. Dey, S. Chakraborty, Phys. Rev. D 105 124051 (2022)
\bibitem{24} S. Chakraborty, K. Chakravarti, S. Bose, S. SenGupta, Phys. Rev. D 97, 104053 (2018)
\bibitem{25} Q. Tan, Wen-Di Guoab, Yu-Xiao Liu, Phys. Rev. D 106, 044038 (2022)
\bibitem{27} P. H. C. Siqueira, M. Richartz, Phys. Rev. D 106, 024046 (2022)
\bibitem{28} T. Torres, S. Patrick, M. Richartz,  S. Weinfurtner, Phys. Rev. Lett. 125, 011301 (2020),
\bibitem{29} T. Assumpcao, V. Cardoso, A. Ishibashi, M. Richartz,  M. Zilhao, Phys. Rev. D 98, 064036 (2018)
\bibitem{30} M. Richartz, Phys. Rev. D 93, 064062 (2016)
\bibitem{31} Wei-Liang Qian , K. Lin, Xiao-Mei Kuang, B. Wang Rui-Hong Yue, Eur. Phys. J. C. 82, 188 (2022),
\bibitem{32} W. Yao, S. Chen, and J. Jing, Phys. Rev. D 83, 124018 (2011)
\bibitem{33} M. Okyay, A. $\ddot{O}$vg $\ddot{u}$n, JCAP 01, 009 (2022)
\bibitem{34} D. Liu, Y. Yang, S. Wu, Y. Xing, Z. Xu,  Z.-W. Long, Phys. Rev. D 104, 104042 (2021)
\bibitem{35} G. Guo, P. Wang, H. Wu, H. Yang, JHEP 06, 060 (2022), arXiv:2112.14133 [gr-qc]
\bibitem{36} M. S. Churilova, R. A. Konoplya,  A. Zhidenko, Phys. Lett. B 802, 135207 (2020)
\bibitem{37} V. Cardoso, S. Hopper, C. F. B. Macedo, C. Palenzuela, P. Pani, Phys. Rev. D 94, 084031 (2016)
\bibitem{38} V. Cardoso, V. F. Foit,  M. Kleban, JCAP 08, 006 (2019)
\bibitem{39} M. R. Correia, V. Cardoso, Phys. Rev. D 97, 084030 (2018)
\bibitem{40} R. A. Konoplya, A. Zhidenko, EPL 138, 49001 (2022)
\bibitem{41} M. S. Churilova, R. A. Konoplya, Z. Stuchlik, A. Zhidenko, JCAP 10, 010 (2021)
\bibitem{42} K. A. Bronnikov, R. A. Konoplya, Phys. Rev. D 101, 064004 (2020)
\bibitem{43} R. A. Konoplya, Z. Stuchlk,  A. Zhidenko, Phys. Rev. D 99, 024007 (2019)
\bibitem{44} V. F. Foit, M. Kleban, Class. Quant. Grav. 36, 035006 (2019)
\bibitem{45} Yu-Tong Wang, Zhi-Peng, J. Zhang, Shuang-Yong Zhou, Yun-Song Piao, Eur. Phys. J. C 78, 482
 (2018)
\bibitem{47} P. Pani,  V. Ferrari, Class. Quant. Grav. 35, 15LT01 (2018)
\bibitem{48} A. Testa, P. Pani, Phys. Rev. D 98, 044018 (2018)
\bibitem{49} E. Maggio, A. Testa, S. Bhagwat,  P. Pani, Phys. Rev. D 100, 064056 (2019)
\bibitem{50} N. Oshita and N. Afshordi, Phys. Rev. D 99, 044002 (2019)
\bibitem{jha} Sohan Kumar Jha: Eur. Phys. J. C (2023) 83:952.
\bibitem{jha1} Sohan Kumar Jha: Eur. Phys. J. Plus (2023) 138:757.
\bibitem{jha2} Himangshu Barman et al.: JCAP05(2024) 019.
\bibitem{jha3} Sohan Kumar Jha: arXiv:2404.15808 [gr-qc].
\bibitem{jha4} Ahmad Al-Badawi, Sohan Kumar Jha 2024 Commun. Theor. Phys. 76 095403.
\bibitem{HAWKING}S. W. Hawking, Commun. Math. Phys. 43, 199 (1975b), [Erratum: Commun.Math.Phys. 46, 206 (1976)]
\bibitem{HH} H. Hassanabadi et al., “Effects of a new extended uncertainty principle on Schwarzschild and Reissner–Nordstr¨om black holes thermodynamics,” Int. J. Mod. Phys. A 36, 2150036 (2021).
\bibitem{SH} S. Hassanabadi et  al., “Thermodynamics of the Schwarzschild and Reissner–Nordstr¨om black holes under higher-order generalized uncertainty principle,” Eur. Phys. J. Plus 136, 918 (2021),
arXiv:2110.01363 [gr-qc].
\bibitem{HC} Hao Chen et  al., “Thermodynamics of the Reissner-Nordstr¨om black hole with quintessence matter on the EGUP framework,” Phys. Lett. B 827, 136994 (2022).
\bibitem{BEK} Black hole thermodynamics, Physics Today 24, Bekenstein 1980.
\bibitem{KEIF} Classical and Quantum black holes by Keifer 1999.
\bibitem{SW} Shao-Wen Wei Zhang, Yu-Peng and Yu-Xiao Liu, “Topological approach to derive the global Hawking temperature of (massive) BTZ black hole.” Physics Lett. B 810 (2020).
\bibitem{ALI} Ali $\ddot{O}$vg$\ddot{u}$n and Izzat Sakalli, “Hawking radiation via gaussbonnet theorem,” Ann. of Phys. 413, 168071 (2020).
\bibitem{SI} S. I Kruglov, “Magnetically charged black hole in framework of nonlinear electrodynamics model.” Int. J. of Mod. Phys. A 33 (2018).
\bibitem{qn32} Creek, S., Efthimiou, O.; Kanti, P. and Tamvakis, K. Phys. Rev. D, 76, 104013 (2007).

\bibitem{qn36} Shankaranarayanan, S. Phys. Rev. D, 67, 084026 (2003).
\bibitem{qn37} Boonserm, P. and Visser, M. . Ann. Phys., 325, 1328--1339 (2010).

\bibitem{qn38} Kanzi, S., Sakall\i , I. Nucl. Phys. B, 946, 114703 (2019).

\bibitem{qn39} Al-Badawi, A., Sakall\i , I. and Kanzi, S. Ann. Phys. 2020, 412, 168026.

\bibitem{qn40} Al-Badawi, A., Kanzi, S. and Sakall\i , I. Eur. Phys. J. Plus 2020, 135,
219.

\bibitem{qn41} P. Boonserm, T. Ngampitipan and P. Wongjun, Eur. Phys. J. C 79, 330 (2019), arXiv:1902.05215 [gr-qc].

\bibitem{qn42} M. Visser, Phys. Rev. A 59, 427 (1999), arXiv:quant-ph/9901030.

\bibitem{qn43} C. V. Vishveshwara, Nature 227, 936 (1970).

\bibitem{qn44} K. D. Kokkotas and B. G. Schmidt, Living Rev. Rel. 2, 2 (1999),
arXiv:gr-qc/9909058.

\bibitem{qn45} H.-P. Nollert, Class. Quant. Grav. 16, R159 (1999).

\bibitem{qn46} V. Cardoso and P. Pani, Living Rev. Rel. 22, 4 (2019), arXiv:1904.05363
[gr-qc].

\bibitem{GB}Matt Visser, “Some general bounds for one-dimensional scattering,” Phys. Rev. A 59, 427–438 (1999).
\bibitem{GB1}Petarpa Boonserm and Matt Visser, “Bounding the bogoliubov coefficients,” Annals of Physics 323, 2779–2798 (2008).
\bibitem{GB2}Boonserm. P, “Rigorous bounds on transmission, reflection and bogoliubov coefficients,” Ph.D. thesis, Victoria Uni. Wellington (2009).
\bibitem{WJ} W. Javed, I. Hussain, and A. O¨ vgu¨n, “”Weak deflection angle of KazakovSolodukhin black hole in plasma medium using GaussBonnet theorem and its greybody bonding.” Eur. Phys. J. Plus 137 (2022).
\bibitem{keeton} C. R. Keeton, C. S. Kochanek, E. E. Falco, The Optical Properties of Gravitational Lens
Galaxies as a Probe of Galaxy Structure and Evolution, Astrophys. J. 509, 561-578
(1998), arXiv:astro-ph/9708161.
\bibitem{eiroa} E. F. Eiroa, G. E. Romero, D. F. Torres, Reissner-Nordstrom black hole lensing, Phys.
Rev. D 66, 024010 (2002), arXiv:gr-qc/0203049.
\bibitem{sharif} M. Sharif, S. Iftikhar, Dynamics of scalar thin-shell for a class of regular black holes,
Astrophys Space Sci 356, 89–101 (2015), arXiv:1511.02762 [physics.gen-ph].
\bibitem{zakharov} A. F. Zakharov, The black hole at the Galactic Center: Observations and models, Int. J.
Mod. Phys. D 27, no.06, 1841009 (2018), arXiv:1801.09920 [gr-qc].
\bibitem{virbhadra} K. S. Virbhadra, D. Narasimha, S. M. Chitre, Role of the scalar field in gravitational
lensing, Astron. Astrophys. 337, 1-8 (1998), arXiv:astro-ph/9801174.
\bibitem{virbhadra1} K. S. Virbhadra, G. F. R. Ellis, Gravitational lensing by naked singularities, Phys. Rev.
D 65, 103004 (2002).
\bibitem{rahaman} F. Rahaman, M. Kalam, S. Chakraborty, Gravitational lensing by stable C-field
wormhole, Chin. J. Phys. 45, 518 (2007), arXiv:0705.0740 [gr-qc].
\bibitem{pkf} P. K. F. Kuhfittig, Gravitational lensing of wormholes in the galactic halo region, Eur.
Phys. J. C 74, no.99, 2818 (2014), arXiv:1311.2274 [gr-qc].
\bibitem{manna} T. Manna, F. Rahaman, S. Molla, J. Bhadra, H. H. Shah, Weak Deflection Angle and
Greybody Bound of Magnetized Regular Black Hole, Gen. Rel. Grav. 50, no.5, 54 (2018).
\bibitem{kumar} R. Kumar, S. G. Ghosh, A. Wang, Gravitational deflection of light and shadow cast by
rotating Kalb-Ramond black holes, Phys. Rev. D 101, no.10, 104001 (2020),
arXiv:2001.00460 [gr-qc].
\bibitem{shaikh} R. Shaikh, P. Kocherlakota, R. Narayan, P. S. Joshi, Shadows of spherically symmetric
black holes and naked singularities, Mon. Not. Roy. Astron. Soc. 482, no.1, 52-64 (2019),
arXiv:1802.08060 [astro-ph.HE].
\bibitem{islam} S. U. Islam, R. Kumar, S. G. Ghosh, Gravitational lensing by black holes in the 4D
Einstein-Gauss-Bonnet gravity, JCAP 09, 030 (2020), arXiv:2004.01038 [gr-qc].
\bibitem{411}Zonghai Li and Ali A.~Ovg{\"u}n, “Finite-distance gravitational deflection of massive particles by a kerr-like black hole in the bumblebee gravity model,” Phys. Rev. D 101, 024040 (2020)
\bibitem{421} Zonghai Li, Guodong Zhang, and Ali A.~Ovg{\"u}n, “Circular orbit of a particle and weak gravitational lensing,” Phys. Rev. D 101, 124058 (2020).
\bibitem{431} J. H. Oort, “The force exerted by the stellar system in the direction perpendicular to the galactic plane and some related problems,” Astron. Inst. Netherlands 6, 249 (1932).
\bibitem{441} Fritz. Zwicky, “On the masses of nebulae and of clusters of nebulae.” The Astrophys. J. 86, 217 (1937).
\bibitem{451} J. L. Feng, “Dark Matter Candidates from Particle Physics and Methods of Detection.” The Astrophys. J. Supple. Ser. 48, 495–545 (2010).
\bibitem{461} N. Jarosik et al., Astrophys. J., Suppl. Ser. 192, 14 (2011).
\bibitem{471} A. A.~Ovg{\"u}n, “Deflection angle of photons through dark matter by black holes and wormholes using gaussbonnet theorem,” Universe 5, 115 (2019).
\bibitem{481} Reggie C. Pantig and A.~Ovg{\"u}n, “Dark matter effect on the weak deflection angle by black holes at the center of Milky Way and M87 galaxies,” Eur. Phys. J. C 82, 391 (2022), arXiv:2201.03365 [gr-qc].
\bibitem{491} Reggie C. Pantig and Emmanuel T. Rodulfo, “Weak deflection angle of a dirty black hole,” Chin. J. Phys. 66, 691–702 (2020).
\bibitem{501} Reggie C. Pantig and A.~Ovg{\"u}n, “Black hole in quantum wave dark matter,” Fortsch. Phys. 2022, 2200164 (2022), arXiv:2210.00523 [gr-qc].
\bibitem{511}Reggie C. Pantig and A.~Ovg{\"u}n, “Dehnen halo effect on a black hole in an ultra-faint dwarf galaxy,” JCAP 08, 056 (2022), arXiv:2202.07404 [astro-ph.GA].
\bibitem{GW} G. W. Gibbons and M. C. Werner, Class. Quant. Grav. 25, 235009 (2008).
\bibitem{WERNER} M. C. Werner, Gen. Rel. Grav. 44, 3047 (2012).
\bibitem{ISHIHARA1} A. Ishihara, Y. Suzuki, T. Ono, T. Kitamura, H. Asada. Phys. Rev., D94(8):084015, (2016).
\bibitem{ISHIHARA2} A. Ishihara, Y. Suzuki, T. Ono and H. Asada, Phys. Rev. D 95, 044017 (2017).
\bibitem{ONO1}T. Ono, A. Ishihara and H. Asada, Phys. Rev. D 96, 104037 (2017).
\bibitem{ONO2} T. Ono, A. Ishihara, H. Asada: Phys. Rev., D96(10):104037, (2017).
\bibitem{ONO3} T. Ono, A. Ishihara, H. Asada. Phys. Rev. D98(4):044047, (2018).
\bibitem{CRISNEJO} G. Crisnejo, E. Gallo and K. Jusufi, Phys. Rev. D100, 104045 (2019).
\bibitem{M871}Kazunori Akiyama et al. First Sagittarius $A^*$ Event Horizon Telescope Results. VI. Testing the Black Hole Metric. Astrophys. J. Lett., 930(2):L17, 2022.
\bibitem{M872}Prashant Kocherlakota et al. Constraints on black-hole charges with the 2017 EHT observations of $M87^*$. Phys. Rev. D, 103(10):104047, 2021.
\bibitem{keck} Tuan Do et al. Relativistic redshift of the star S0-2 orbiting the Galactic center supermassive black hole. Science, 365(6454):664-668, 2019.
\bibitem{vlti1} R. Abuter et al. Mass distribution in the Galactic Center based on interferometric astrometry of multiple stellar orbits. Astron. Astrophys., 657:L12, 2022.
\bibitem{vlti2} R. Abuter et al. Detection of the Schwarzschild precession in the orbit of the star S2 near the Galactic centre massive black hole. Astron. Astrophys., 636:L5, 2020.
\bibitem{Iyer} S. Iyer, C.M. Will, Phys. Rev. D 35, 3621 (1987)
\bibitem{Konoplya}  R.A. Konoplya, Phys. Rev. D 68 024018 (2003)
\bibitem{yg2017}Y.-G. Miao and Z.-M. Xu, Hawking Radiation of Five-Dimensional Charged Black Holes with Scalar
Fields, Phys. Lett. B 772, 542 (2017).
\bibitem{fg2016} F. Gray, S. Schuster, A. Van–Brunt, and M. Visser, The Hawking Cascade from a Black Hole Is
Extremely Sparse, Class. Quantum Grav. 33, 115003 (2016).
\bibitem{CARMO}M. P. Do Carmo, Differential Geometry of Curves and Surfaces, (Prentice-Hall, New Jersey, 1976).
\bibitem{del} Kazunori Akiyama et al. First Sagittarius $A^*$ Event Horizon Telescope Results. II. EHT and Multiwavelength Observations, Data Processing, and Calibration. Astrophys. J. Lett., 930(2):L13, 2022.
\end{thebibliography}
\end{document}